# Transferable screened range-separated hybrid functionals for electronic and optical properties of van der Waals materials

María Camarasa-Gómez,[1] Ashwin Ramasubramaniam,[2, 3] Jeffrey B. Neaton,[4, 5, 6] and Leeor Kronik[1]

[1] Department of Molecular Chemistry and Materials Science,
Weizmann Institute of Science, Rehovoth 7610001, Israel
[2] Department of Mechanical and Industrial Engineering,
University of Massachusetts Amherst, Amherst MA 01003, USA
[3] Materials Science and Engineering Graduate Program,
University of Massachusetts, Amherst, Amherst MA 01003, USA
[4] Department of Physics, University of California, Berkeley, CA 94720, USA
[5] Materials Sciences Division, Lawrence Berkeley National Laboratory, Berkeley, CA 94720, USA
[6] Kavli Energy NanoSciences Institute at Berkeley,
University of California, Berkeley, CA 94720, USA

(Dated: May 22, 2023)

The accurate description of electronic properties and optical absorption spectra is a long-standing challenge for density functional theory. Recently, the introduction of screened range-separated hybrid (SRSH) functionals for solid-state materials has allowed for the calculation of fundamental band gaps and optical absorption spectra that are in very good agreement with many-body perturbation theory. However, since solid-state SRSH functionals are typically tuned to reproduce the properties of bulk phases, their transferability to low-dimensional structures, which experience substantially different screening than in the bulk, remains an open question. In this work, we explore the transferability of SRSH functionals to several prototypical van der Waals materials, including transition-metal sulfides and selenides, indium selenide, black phosphorus, and hexagonal boron nitride. Considering the bulk and a monolayer of these materials as limiting cases, we show that the parameters of the SRSH functional can be determined systematically, using only the band-edge quasiparticle energies of these extremal structural phases as fitting targets. The resulting SRSH functionals can describe both electronic bandstructures and optical absorption spectra with accuracy comparable to more demanding ab initio many-body perturbation theory (GW and Bethe-Salpeter equation) approaches. Selected examples also demonstrate that the SRSH parameters, obtained from the bulk and monolayer reference structures, display good accuracy for bandstructures and optical spectra of bilayers, indicating a degree of transferability that is independent of the fitting procedure.

## I. INTRODUCTION

Van der Waals (vdW) layered materials have been in the spotlight for almost two decades [1–3], attracting an enormous amount of attention since the experimental isolation of graphene in 2004 [4, 5]. These materials present an inherently wide range of structural, electronic, and optical properties, which is vastly enhanced by the possibility of combining layers (e.g., through heterostructuring, as well as relative twisting or sliding) to allow for additional tuning of their physicochemical properties.[1–3, 6–12].

As the space of vdW materials and their derivatives continues to expand, there is an ever growing need for a reliable theoretical description of their electronic and optical properties, particularly bandstructures and optical absorption spectra. Presently, state-of-the-art first-principles calculations of these properties in crystalline materials are based mostly on ab initio many-body perturbation theory (MBPT) [13, 14]. Ab initio MBPT is usually employed in practice by using the GW approximation [15] with input from a (generalized) Kohn-Sham eigensystem to obtain single quasi-particle excitation energies and the Bethe-Salpeter equation (BSE) [16, 17] to calculate neutral excitation energies and optical absorption spectra. Indeed, GW-BSE has been found to be very successful in the intepretation and even prediction of electronic and optical properties in vdW materials (e.g., refs [18–23]). However, GW-BSE calculations are relatively expensive computationally [24–26] and can become prohibitively expensive especially when supercells are called for, e.g. in the calculation of defects or of twisted multilayer structures. Therefore, there remains a need for alternative computational approaches that can provide results with similar accuracy at a substantially lower computational cost.

A first-principles alternative to MBPT is density functional theory (DFT) [27, 28], as well as its extension to excited states, namely time-dependent DFT (TDDFT) [29, 30]. However, it is well known that (TD)DFT with common approximate functionals often fails in the prediction of electronic and optical excitations in solids [14]. One promising recent approach within DFT is that of the tuned screened range-separated hybrid functional (SRSH) [31, 32]. Originally applied primarily to molecular solids [31, 33–35], these functionals have recently been found to be extremely useful in the determination of fundamental gaps and optical spectra of semiconductors and insulators [36–45]. Moreover, recently a non-empirical optimally-tuned SRSH approach that is appli-cable to a general semiconductor or insulator has been found to be quantitatively predictive for a wide range of materials, from narrow-gap semiconductors to wide-gap insulators [39, 40, 46].

A key feature of SRSH functionals is the proper incorporation of dielectric screening in the long-range exchange, which assures the correct asymptotic decay of the Coulomb tail [31–33, 47, 48]. However, in vdW materials the dielectric constant varies with the number of layers between its bulk value and unity – formally, the correct asymptotic limit for screening in a monolayer [49, 50] – introducing a potentially large structure dependence to the parameters of the SRSH functional. Therefore, an open question remains as to whether an SRSH functional that is optimally tuned, say, for the bulk phase of a material, would be transferable to lower-dimensional structures. In this article, we expand our prior work [37] on assessing and addressing this topic to encompass a broad range of semiconducting and insulating vdW materials, exploring both monolayers (2D materials) and bulk phases of Mo- and W-based transition-metal dichalcogenides, indium selenide, black phosphorus, and hBN, as well as bilayers of MoS₂ and hBN. We find that SRSH functionals can be tuned, using only a single quasiparticle energy gap at the band edges of the 2D and bulk phases, to produce bandstructures that are in excellent agreement with GW calculations over the entire Brillouin zone. Furthermore, we demonstrate that time-dependent (TD) SRSH calculations can produce optical absorption spectra for vdW semiconductors that are in very good agreement with spectra from GW-BSE calculations without any input from the latter in the tuning procedure, rendering the TD-SRSH approach truly predictive. We also explore the transferability of the SRSH parameters – derived for 2D and bulk phases – to bilayers of MoS₂ and h-BN, and show that their bandstructures continue to be in good agreement with GW calculations. Further computational details can be found in the Supplementary Information (SI).

## II. THEORY OF THE SRSH FUNCTIONAL

In the SRSH approach, the Coulomb operator is partitioned into short range (SR) and long range (LR) components through the introduction of three parameters, $\alpha$, $\beta$, $\gamma$, as follows [47, 48]:

$$\frac{1}{r} = \frac{\alpha + \beta \, \mathrm{erf}(\gamma r)}{r} + \frac{1 - [\alpha + \beta \, \mathrm{erf}(\gamma r)]}{r}, \qquad (1)$$

where $\mathrm{erf}(\cdot)$ is the error function, $r$ is the inter-electron distance, and $\gamma$ is a range-separation parameter. The first term of Eq. (1) is treated using exact exchange while the second term is treated using a semilocal approximation. The parameter $\alpha$ therefore sets the fraction of exact exchange in the short range, the sum of parameters $\alpha + \beta$ sets the fraction of exact exchange in the long range, and $1/\gamma$ provides a length-scale for the crossover from

short- to long-range behavior, interpolated smoothly by the error function [51]. To enforce a correct asymptotic behaviour of the screened Coulomb operator via appropriate dielectric screening, we impose the condition $\alpha + \beta = 1/\epsilon_\infty$, where $\epsilon_\infty$ is the high-frequency scalar (orientationally-averaged) dielectric constant [33]. Enforcing this limit is essential to capturing excitonic effects in solid-state systems [33, 36, 52]. The above relation fixes the value of $\beta$, given a choice of $\alpha$, in terms of $\epsilon_\infty$, leaving two free parameters: $\alpha$ and $\gamma$. This approach neglects anisotropy and approximates the dielectric constant as a scalar, i.e., $\epsilon_\infty = \mathrm{Tr}[\epsilon_\infty]/3$. For 2D systems, we set $\epsilon_\infty = 1$, which is the correct asymptotic limit of screening in the long range for an isolated 2D system [49, 50].

With these ingredients at hand, the exchange potential of the SRSH functional, derived within generalized Kohn-Sham theory [53–56], is represented by the non-multiplicative potential operator

$$v_x^{\mathrm{SRSH}} = \alpha v_{\mathrm{XX}}^{\mathrm{SR}} + (1-\alpha) v_{\mathrm{SL}}^{\mathrm{SR}} + \frac{1}{\epsilon_\infty} v_{\mathrm{XX}}^{\mathrm{LR}} + \left(1 - \frac{1}{\epsilon_\infty}\right) v_{\mathrm{SL}}^{\mathrm{LR}}, \tag{2}$$

where the subscripts '$x$', 'XX', and 'SL' denote exchange, exact (Fock) exchange, and semi-local exchange, respectively.

In this work, we follow the approach of Ref. [37] to obtain the parameters $\alpha$, $\gamma$, and $\epsilon_\infty$ that fully determine the SRSH functional for a given material. First, the scalar dielectric constant, $\epsilon_\infty$, for bulk phases is determined non-empirically using the random phase approximation [57] (RPA) that includes local-field effects at the Hartree level of a (semi-)local functional. Other approaches are equally valid, but we choose to use the RPA to maintain consistency between the treatment of the dielectric response in the DFT and the GW calculations. The calculated values of $\epsilon_\infty$ for the bulk phases of various materials studied here are listed in Table I. For monolayers and bilayers, we use a value of $\epsilon_\infty = 1$, as discussed above. To determine suitable values of $\alpha$ and $\gamma$ for each material, we perform GW calculations to determine the quasiparticle gaps for the bulk and monolayer structures. For monolayers, these quasiparticle gaps are extrapolated to the limit of infinite vacuum (see SI). We then perform a sweep over the $\alpha - \gamma$ parameter space and calculate the corresponding SRSH band gaps for pairs of values $(\alpha, \gamma)$. We quantify the error in the SRSH calculation, relative to the reference GW result, by the difference between the quasiparticle and SRSH band gaps at a particular $k$-point, $\Delta E_g = E_g^{\mathrm{GW}} - E_g^{\mathrm{SRSH}}$. We find that it is generally sufficient to ensure that $\Delta E_g = 0$ at just one high-symmetry $k$-point, which we pick to correspond to the smallest direct band gap.

As noted in prior work [36–38, 58], the choice of $\alpha$ and $\gamma$ is not unique for a particular material (or phase) and the tuning procedure outlined above generally leads to a continuum of values that lie on a "zero-crossing" line, $\Delta E_g = 0$, of the $\Delta E_g(\alpha, \gamma)$ surface. Given two phases – the bulk and monolayer – the intersection of their individual zero-crossing lines leads to a unique set of parameters, $(\alpha^*, \gamma^*)$, that is simultaneously optimal for both phases. It is this optimal pair that is finally used for computing electronic and optical properties of the various materials.

Using this procedure we determine transferable pairs, $(\alpha^*, \gamma^*)$, for various vdW materials and compare their electronic bandstructures against those obtained from GW calculations. We also report optical absorption spectra obtained from linear response [59] TD-SRSH calculations [31, 35, 36], and compare the results against GW-BSE calculations within the Tamm-Dancoff approximation [60].

## III. RESULTS AND DISCUSSION

In this section, we apply the methodology described in Section II to representative vdW materials: these include transition-metal dichalcogenides (TMDCs) $WS_2$, $WSe_2$ and $MoSe_2$; black phosphorus (BP); and InSe. These materials range from narrow- to small-gap semiconductors.

### A. Transition-Metal Dichalcogenides: $WS_2$, $WSe_2$, and $MoSe_2$

$WS_2$, $WSe_2$, and $MoSe_2$ are semiconductors that crystallize in the trigonal prismatic 2H phase (space group P3̄m1) in their ground state. Figure 1 (a-c) displays contour plots of the error in the band gap, $\Delta E_g$, for bulk and monolayers of the three materials. Interestingly, the errors in the band gap present similar trends across this group of materials: the bulk phases exhibit a very small degree of acceptable variation in the fraction of short-range exact exchange, $\alpha$, whereas this parameter can vary more widely for monolayers. It is also clear from these figures that optimizing the SRSH for just one phase can lead to rather large errors for the other phase. For example, selecting acceptable values of $(\alpha, \gamma)$ for bulk $WS_2$ at the extremes of the $\alpha - \gamma$ plot (Fig. 1a) leads to large errors in the predicted band gap of the monolayer, ranging from $-0.122$ eV for $(\alpha, \gamma) = (0.102, 0.010\,\text{Å}^{-1})$ to 2.33 eV for $(\alpha, \gamma) = (0.116, 0.248\,\text{Å}^{-1})$. Conversely, optimizing the SRSH purely for monolayer $WS_2$ results in errors in the bulk band gap ranging from 0.030 eV for $(\alpha, \gamma) = (0.011, 0.044\,\text{Å}^{-1})$ to 0.049 eV for $(\alpha, \gamma) = (0.113, 0.010\,\text{Å}^{-1})$. The point of intersection of the zero-crossings of the gap deviation surfaces $(\alpha^*, \gamma^*) = (0.102, 0.019\,\text{Å}^{-1})$, simultaneously renders the error in the band gap zero for both phases. Similar behavior is observed for $WSe_2$ and $MoSe_2$.

Table I displays the optimal parameters, $\alpha^*$ and $\gamma^*$, along with the RPA dielectric constants, $\epsilon_\infty$. Figure 1 displays the corresponding SRSH bandstructures along with the corresponding GW bandstructures. As the SRSH functionals were tuned to reproduce GW band gaps extrapolated to infinite interlayer separation and infinite k-point sampling (see SI), the outcome of a particular un-extrapolated GW calculation will always differ to some extent from the SRSH result. For example, for $WS_2$, the un-extrapolated GW band gap for the monolayer (at the $K$ point) is 30 meV larger than the SRSH band gap and the un-extrapolated GW band gap for the bulk (at the $K$ point) is 40 meV larger than its SRSH counterpart. Similar differences ($\sim 20$ meV) are found for $MoSe_2$ and $WSe_2$. The above small differences notwithstanding, Figure 1 (d-f) shows very good agreement between the GW and SRSH bandstructures for all three, especially at the band edges. Qualitatively, the deviations are somewhat larger for the bulk than for the monolayers, particularly for the selenides. These deviations also become more apparent deeper into the valence or conduction band, which is generally expected when using SRSH eigenvalues as approximate quasi-particle excitation energies [54, 61, 62]. These deeper bands, however, are less relevant to electronics applications or to the low-energy optical absorption spectrum. The mean absolute deviation between the GW and SRSH results for the top-most valence band and bottom-most conduction band, over all $k$ points, is 0.060 eV, 0.063 eV and 0.103 eV for monolayer $WS_2$, $WSe_2$ and $MoSe_2$, respectively, and 0.059 eV, 0.074 eV and 0.098 eV for the bulk.

We note that the parameter $\gamma$ is relatively small and similar to the value for $MoS_2$ reported in Ref. 37. As a consequence, one might be tempted to conclude that the SRSH functional behaves almost as the corresponding limit of a global hybrid [63]. However, in the $\gamma \to 0$ limit, $\beta$ would be irrelevant and the exchange would be asymptotically screened by $1/\alpha$ instead of $\epsilon_\infty$, with consequences for predicted exciton binding energies. For example, in Chen et al. [58], the fraction of exact exchange was tuned to fulfill the ionization potential theorem in a system with a defect. It was concluded in that work that using a global hybrid that is tuned to the band gap at only one $k$-point may lead to inaccurate electronic structure predictions. This underscores the importance of using a range-separated hybrid rather than a global one.

Next, we consider optical absorption spectra for the same materials, as shown in Figure 2. These calculations were performed without spin-orbit coupling, primarily due to the computational cost of the reference GW-BSE calculations. Insets provide corresponding TD-SRSH spectra for monolayers that do include spin-orbit coupling. Recalling that SRSH parameters are only tuned to reproduce the GW band gap at a single $k$-point, any further calculations with the same parameters are true tests of the predictive capability of the functional. As seen in Figure 2, the overall agreement between the GW-BSE and TD-SRSH spectra is highly satisfactory, especially for the low-energy part of the spectrum. At higher energies ($\gtrsim 2.5$ eV), some disagreement becomes more apparent, most likely due to the above-noted larger deviations between higher and lower lying SRSH and GW eigenvalues. Nonetheless, the agreement between GW-BSE and TD-SRSH spectra is particularly good for the bulk, and the neglecting of the anisotropy of the dielectric constant does not seem to have introduced qualitative failures in the monolayer calculations. For the latter, the positions of the low-energy peaks are generally in good agreement between TD-SRSH and GW-BSE (deviations smaller than 0.1 eV) whereas the discrepancy in peak heights is more apparent.

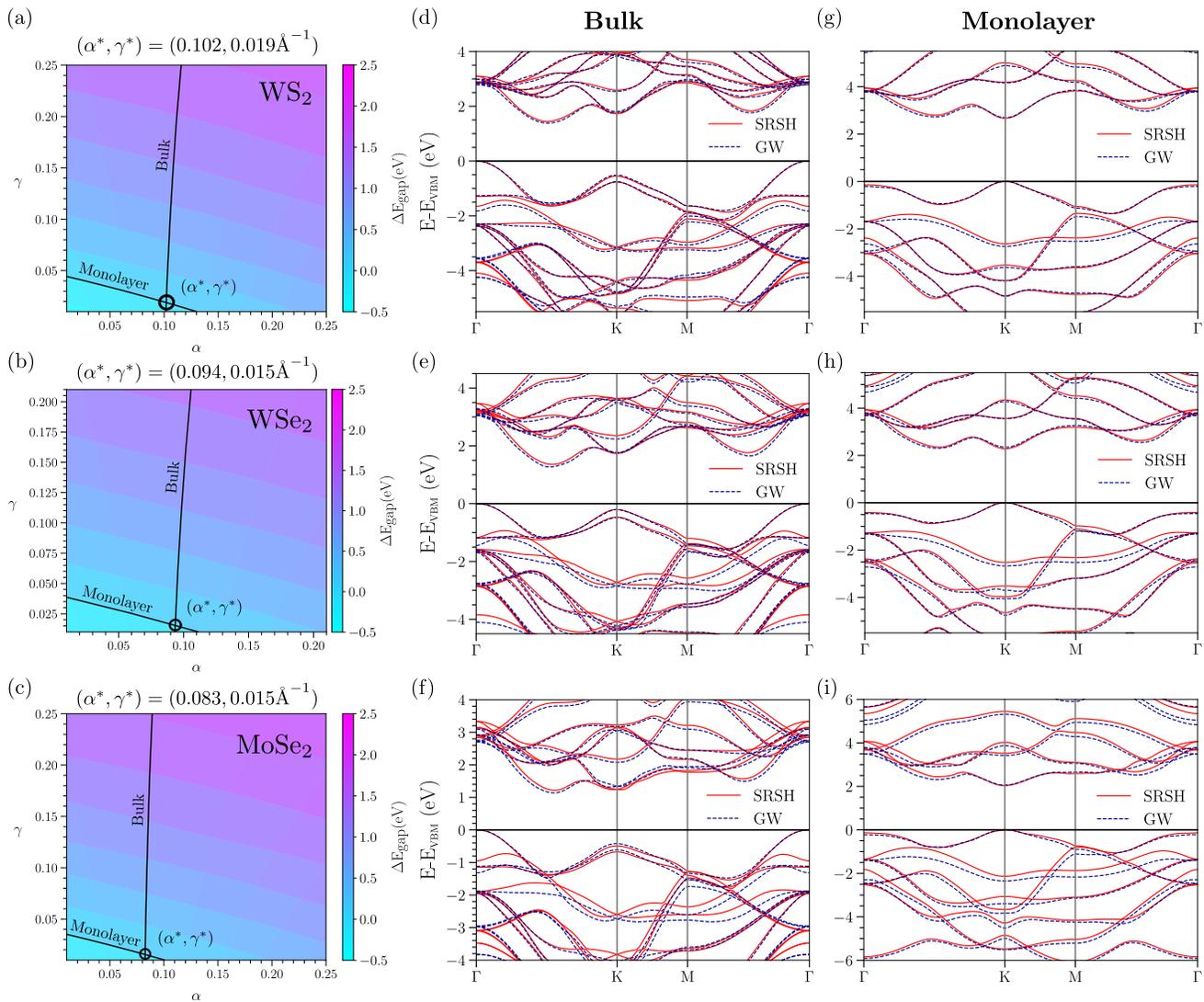

FIG. 1. (a)-(c) Contour maps of the gap deviation, $\Delta E_g$, for WS$_2$, WSe$_2$, and MoSe$_2$. The solid black lines represent the values for which $\Delta E_g = 0$ for bulk and monolayer structures, and the intersection of the two lines yields a unique set of values $(\alpha^*, \gamma^*)$ that are transferable between the bulk and monolayer. (d)-(f) Bandstructures for bulk and (g)-(i) bandstructures for monolayers of WS$_2$, WSe$_2$, and MoSe$_2$ from SRSH (solid lines) and G$_0$W$_0$@PBE (dashed lines). $\gamma$ has units of Å$^{-1}$.

Upon inclusion of spin-orbit coupling in the SRSH monolayer calculations, we observe the appearance of the characteristic A and B excitonic peaks of TMDC monolayers. For WS$_2$, the experimentally measured A and B peaks are at 2.12 eV and 2.5 eV [64], respectively, which compares excellently to the TD-SRSH peaks located at 2.05 eV and 2.38 eV. Likewise, for WSe$_2$, the experimen-tal values are 1.74 eV and 2.16 eV [65], compared to TD-SRSH values at 1.70 eV and 1.98 eV, which represents a slightly larger, but still small, deviation. Lastly, for MoSe$_2$, the experimentally measured A and B peaks are at 1.64 eV and 1.83 eV [66], respectively, while the TD-SRSH peaks are located at 1.63 eV and 1.85 eV.

### B. Black Phosphorus and Phosphorene

Among the allotropes of phosphorus, black phosphorus (BP) is one of the most stable forms under ambient conditions [67]. In its ground state, BP is composed of puckered monolayers arranged in an AB stacked structure (space group 64, Cmce) and is very sensitive to changes in pressure [68, 69]. It is also characterized by strong in-plane anisotropy that provides an opportunity for exploiting its orientation-dependent optoelectronic properties in a variety of applications [69–71]. The band gap of BP varies from about 0.3 eV in the bulk to around 2.0 eV for the monolayer (phosphorene), covering much of the range between semiconducting TMDCs and gapless graphene [69, 70, 72].

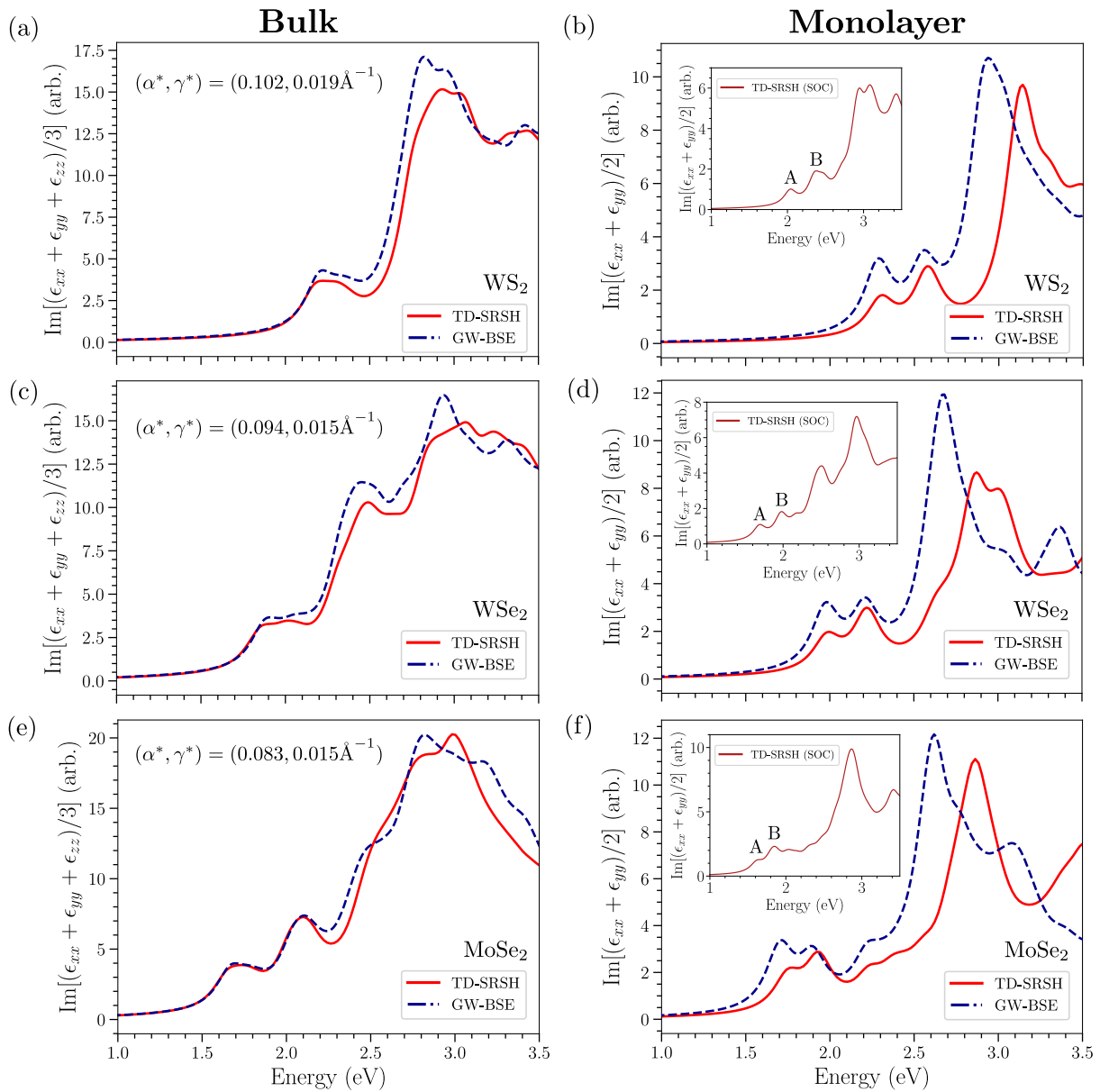

FIG. 2. Optical absorption spectra calculated, without spin-orbit coupling, using TD-SRSH (red solid line) and GW-BSE (blue dashed line). Rows corresponds to WS$_2$, WSe$_2$, and MoSe$_2$. Results for the bulk and for the monolayer are given in the left and right columns, respectively. Insets: corresponding monolayer SRSH calculations that include spin-orbit coupling, in which the A and B peaks represent excitons of the TMDCs. See Table I for specific parameters.

Because BP is a narrow-gap semiconductor, semilocal functionals such as PBE [73] predict an incorrect metallic ground state for this material, creating a qualitatively incorrect starting point for "single-shot" GW calculations [74–76]. Therefore, here we employed the

HSE06 short-range hybrid functional [77, 78] to produce a gapped (0.317 eV direct gap at Γ [79]) starting point for the GW calculation. The use of hybrid functionals as a starting point for perturbative GW calculations is a topic of ongoing research (e.g., [46, 76, 80–83]). In the present case, this approach yielded a G$_0$W$_0$@HSE extrapolated band gap of 0.56 eV, in good agreement with Refs. [83, 84].

Figure 3(a,b) displays surfaces and contour plots of the error, $\Delta E_g$, in the SRSH band gap relative to the GW (G$_0$W$_0$@HSE) fitting target (see SI for details). The zero crossing lines of the $\Delta E_g$ surfaces for BP and

TABLE I. Brillouin zone sampling, tuned SRSH parameters ($\alpha^*, \gamma^*$), average inverse macroscopic dielectric constant ($\epsilon_\infty^{-1}$), GW band gap ($E^{\mathrm{GW}}$), GW-BSE optical gap ($E_{\mathrm{opt}}^{\mathrm{GW-BSE}}$), SRSH band gap ($E^{\mathrm{SRSH}}$, fitted to an extrapolated GW quasiparticle band gap), and TD-SRSH optical gap ($E_{\mathrm{opt}}^{\mathrm{TD-SRSH}}$), for the various materials studied in this article. Additional computational details are given in the SI. Band gaps and optical gaps are calculated at the K point for the TMDC materials ($WS_2$, $WSe_2$, $MoSe_2$) and at the $\Gamma$ point for black phosphorus (BP) and InSe.

| Material | Phase | k-grid | $\alpha^*$ | $\gamma^*(\text{Å}^{-1})$ | $\epsilon_\infty^{-1}$ | $E^{\mathrm{GW}}$ [eV] | $E_{\mathrm{opt}}^{\mathrm{GW-BSE}}$ [eV] | $E^{\mathrm{SRSH}}$ [eV] | $E_{\mathrm{opt}}^{\mathrm{TD-SRSH}}$ [eV] |
|---|---|---|---|---|---|---|---|---|---|
| **$WS_2$** | Bulk | $12 \times 12 \times 4$ | 0.102 | 0.019 | 0.093 | 2.32 | 2.22 | 2.29 | 2.21 |
| | 1L | $18 \times 18 \times 1$ | | | 1.0 | 2.70 | 2.29 | 2.66 | 2.32 |
| **$WSe_2$** | Bulk | $12 \times 12 \times 4$ | 0.094 | 0.015 | 0.084 | 1.97 | 1.91 | 1.94 | 1.89 |
| | 1L | $18 \times 18 \times 1$ | | | 1.0 | 2.34 | 1.98 | 2.28 | 1.99 |
| **$MoSe_2$** | Bulk | $12 \times 12 \times 4$ | 0.083 | 0.015 | 0.079 | 1.76 | 1.70 | 1.74 | 1.74 |
| | 1L | $18 \times 18 \times 1$ | | | 1.0 | 2.06 | 1.72 | 2.04 | 1.78 |
| **BP** | Bulk | $8 \times 8 \times 4$ | 0.170 | 0.035 | 0.090 | 0.49 | 0.40 | 0.56 | 0.32 |
| | 1L | $15 \times 15 \times 1$ | | | 1.0 | 1.89 | 1.34 | 1.95 | 1.38 |
| **InSe** | Bulk | $9 \times 9 \times 4$ | 0.149 | 0.021 | 0.121 | 1.15 | 1.05 | 1.21 | 1.10 |
| | 1L | $15 \times 15 \times 1$ | | | 1.0 | 2.90 | 2.67 | 2.87 | 2.76 |

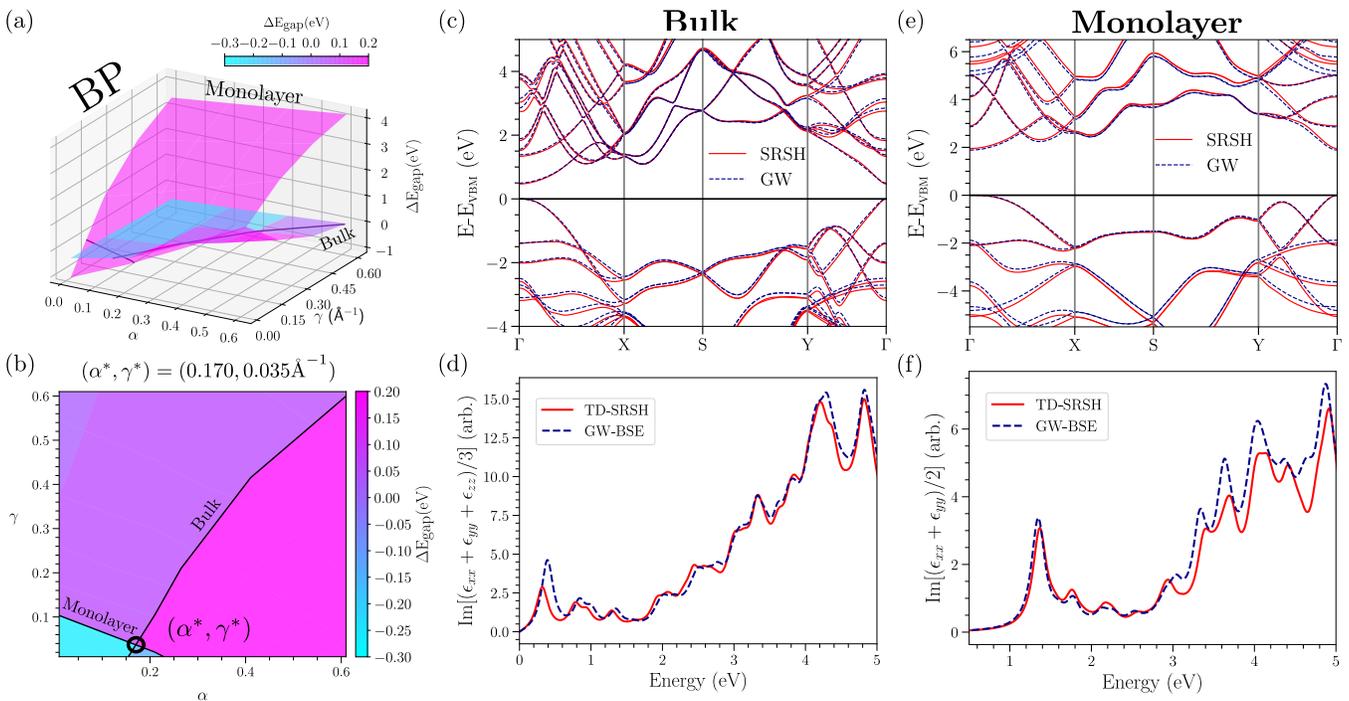

FIG. 3. (a) Contour maps of the gap deviation, $\Delta E_g$, and (b) its projection over the $\alpha - \gamma$ plane for black phosphorus. The solid black lines represent the values for which $\Delta E_g = 0$ for bulk and monolayer structures, and the intersection of the two lines yields a unique set of values ($\alpha^*, \gamma^*$) that are transferable between the bulk and monolayer. (c) Bandstructures for bulk and (e) monolayers of black phosphorus from SRSH (solid lines) and $G_0W_0@PBE$ (dashed lines). (d) Optical absorption spectra for bulk and (f) monolayers of black phosphorus from TD-SRSH (solid red lines) and $G_0W_0$-BSE (dashed-dotted blue lines). $\gamma$ has units of $\text{Å}^{-1}$.

phosphorene intersect at $(\alpha^*, \gamma^*) = (0.170, 0.035\,\text{Å}^{-1})$, which furnishes the optimal set of parameters for further SRSH/TD-SRSH calculations. The complete set of parameters is listed in Table I.

Using these parameters, Figures 3(c,d) and 3(e,f) display corresponding band structures and optical absorption spectra for BP and phosphorene. Once again, owing to the tuning to extrapolated GW band gaps, there is a small deviation between un-extrapolated GW and SRSH values, but beyond that, once again we find excellent agreement between the GW and SRSH bandstructures, both for BP and phosphorene, not just at the band edges but also up to ∼ 2 eV into the valence and conduction bands. Again, the mean absolute deviation for the high-est valence band and lowest conduction band across the entire Brillouin zone is a mere 0.054 eV for the bulk phase and 0.071 eV for the monolayer. For the optical spectra, we find excellent agreement between the two approaches, with deviations in peak positions being at most 60 meV in the low-energy part of the spectrum ($\lesssim 2.5$ eV). The first excitonic peak for phosphorene is located at 1.2 eV, in agreement with a previous BSE study [72]. Here, TD-SRSH and GW-BSE peak heights are also in better agreement than for the TMDCs.

### C. Indium Selenide

As a final example, we consider $\beta$-InSe (space group P6$_3$/mmc). This is a transition-metal monochalcogenide that is part of a larger group of similar materials composed of a Group IIIA element (In, Ga) and a chalcogen (S, Se, Te) [85]. This material exhibits a band gap that changes from 2.87 eV (indirect gap) for a monolayer (theoretical) [86] to 1.20-1.28 eV (direct gap) for the bulk [87–89], as well as high-carrier mobility [85, 90], making it a desirable candidate for optoelectronics [85, 91–93].

Figure 4(a, b) display the surfaces and contour plots of the gap deviation, $\Delta E_g$, in the SRSH band gap relative to the GW fitting targets for the bulk and monolayer structures (see SI). Similar to the TMDCs, we observe that bulk phase exhibits a very small degree of acceptable variation in the fraction of short-range exact exchange, $\alpha$, whereas this parameter can vary more widely for the monolayer. The optimal set of SRSH parameters, $(\alpha^*, \gamma^*) = (0.149, 0.021 \text{ Å}^{-1})$, is obtained from the point of intersection of the zero-crossing lines of the $\Delta E_g$ surfaces for bulk and monolayer InSe. The complete set of parameters for InSe is listed in Table I. Using this tuned SRSH functional, Figure 4(c, e) shows the bandstructures of bulk and monolayer InSe, along with reference bandstructures from GW calculations. Once again, owing to extrapolation the SRSH gaps differ from the GW ones by 60 meV for the bulk and 30 meV for the monolayer. As seen in the figure, the agreement between the SRSH and GW bandstructures is quite satisfactory across the chosen high-symmetry paths, with a mean absolute deviation for the top valence band and the bottom conduction band of 0.163 eV for the bulk and 0.085 eV for the monolayer.

TD-SRSH optical absorption spectra for bulk and monolayer InSe are displayed in Figure 4(d, f), along with reference GW-BSE spectra. The TD-SRSH and GW-BSE spectra are in good agreement below $\sim 3$ eV with the largest error in the energies of the first two peaks being of the order of 0.2 eV, albeit with some differences in the oscillator strength. The agreement between the TD-SRSH and GW-BSE optical spectra is not as good above 3 eV. The remaining discrepancies may be partly due to computational limitations in $k$-point sampling in the GW calculations. The inset of Figure 4(f) displays the TD-SRSH absorption spectrum for the monolayer with the inclusion of spin-orbit coupling; the first excitonic peak (labeled A) appears at 2.6 eV, and agrees well with the value of $\sim 2.57$ eV reported in Ref. [94].

## IV. ASSESSMENT OF SRSH FUNCTIONALS FOR BILAYER MoS$_2$ AND h-BN

Motivated by the promising results of the SRSH/TD-SRSH approach for bulk and monolayer structures, we now seek to understand how well these functionals perform for bilayers of vdW materials. In general, one could expect that as long as the characteristic length scale for switching from short-range exact exchange ($\alpha$) to long-range exact exchange ($1/\epsilon_\infty$), namely $1/\gamma$, is greater than the thickness ($t$) of the bilayer/few-layer slab, the interaction between two charges separated across the slab thickness will be governed largely by the (tuned) short-range exchange. In this scenario, it is reasonable to hypothesize that the SRSH/TD-SRSH formalism ought to retain its accuracy for bilayer/few-layer structures, even when merely employing the simply functional form of the SRSH with asymptotic long-range screening of $\epsilon_\infty = 1$. In the following, we test this hypothesis for bilayer MoS$_2$ and h-BN, bulk and monolayers of which were studied previously in Ref. 37.

### A. MoS$_2$

Tuned SRSH parameters, $(\alpha^*, \gamma^*)$, for bulk and monolayer MoS$_2$ were reported previously in Ref. 37 and are listed in Table II. In principle, one could use these parameters directly to make a *prediction* for bilayer MoS$_2$. It is also possible to re-tune the SRSH using bilayer MoS$_2$ and the bulk as reference structures. To this end, we first perform GW calculations for MoS$_2$ bilayers to determine the reference quasiparticle band gaps (see SI and Table II). We then apply our tuning procedure (Section II) to obtain the error, $\Delta E_g$, for the bilayer as a function of $\alpha$ and $\gamma$. Figure 5(a, b) display the $\Delta E_g$ surfaces and contour plots for monolayer, bilayer, and bulk MoS$_2$. As seen in the figure, the optimal parameters for the bilayer and bulk, labeled $\alpha'$ and $\gamma'$, are not identical to those of the monolayer and bulk ($\alpha^*, \gamma^*$). Specifically, $\alpha'$ and $\alpha^*$ do not vary substantially, as the bulk constrains these values to a rather small window, and the main distinction is manifested in the values of $\gamma'$ and $\gamma^*$. In addition, the $\Delta E_g$ surfaces of the monolayer and bilayer and, consequently, the zero-crossing lines are nearly parallel to each other. This indicates that it is not possible to render $\Delta E_g = 0$ simultaneously for the monolayer and bilayer, though this does not rule out simultaneous minimization of a different metric, an issue we do not explore further here.

Figures 5(c, d) display the SRSH bandstructure and TD-SRSH optical absorption spectrum for bilayer MoS$_2$, using the optimal values of $\alpha'$ and $\gamma'$, along with their

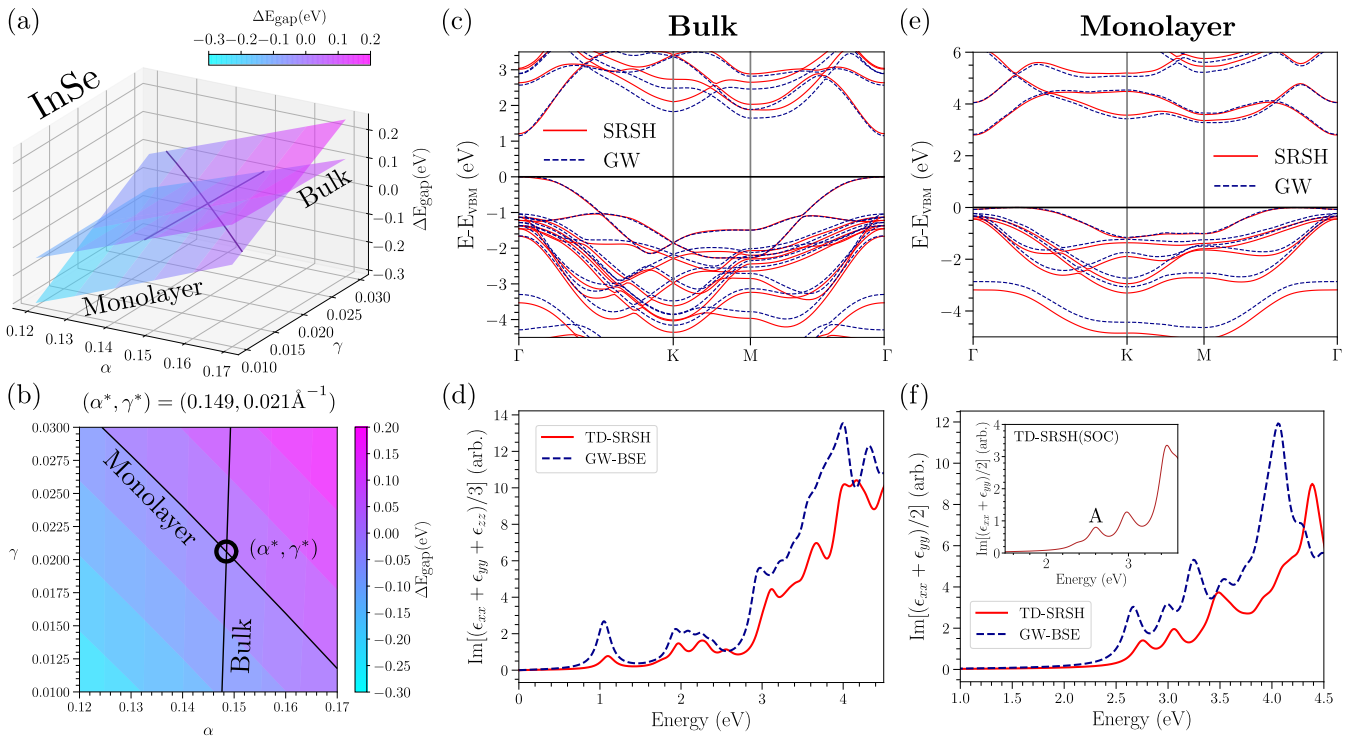

FIG. 4. (a) Contour maps of the gap deviation, $\Delta E_g$, and (b) its projection over the $\alpha - \gamma$ plane for InSe. The solid black lines represent the values for which $\Delta E_g = 0$ for bulk and monolayer structures, and the intersection of the two lines yields a unique set of values $(\alpha^*, \gamma^*)$ that are transferable between the bulk and monolayer. (c) Bandstructures for bulk and (d) monolayers of black phosphorus from SRSH (solid lines) and $G_0W_0$@PBE (dashed lines). (e) Optical absorption spectra for bulk and (f) monolayers of InSe from TD-SRSH (solid red lines) and $G_0W_0$-BSE (dashed-dotted blue lines). $\gamma$ has units of Å$^{-1}$.

GW and GW-BSE counterparts. As before, there is an intrinsic 20 meV extrapolation difference. The overall agreement between the two approaches is excellent: the mean absolute error in the energy eigenvalues, considering the lowermost conduction band and the uppermost valence band, is a mere 0.087 eV. Similarly, we also find good agreement between the GW-BSE and TDSRSH optical spectra (Fig. 5(d)) with differences of less than 0.1 eV in peak positions for the low-energy part of the spectrum ($\lesssim 2.5$ eV). The inset of Figures 5(d) displays the TD-SRSH absorption spectrum with spin-orbit coupling included. We observe the characteristic splitting of the valence band into A and B excitonic peaks at 1.88 eV and 2.23 eV, respectively, that are in excellent agreement with the reported experimental values of 1.91 eV (A peak) and 2.12 eV (B peak) [95].

Returning to the issue of the transferability of the SRSH functional, we sought to understand the implications of modeling the $MoS_2$ bilayer using a functional specifically tuned for the monolayer and bulk (parameters $\alpha^*$ and $\gamma^*$), and conversely, modeling the monolayer using a functional specifically tuned for the bilayer and bulk (parameters $\alpha'$ and $\gamma'$). Figures 6 (a) present the outcome of such a comparison for the bilayer, bulk, and monolayer, with Figures 6 (b) displaying the corresponding TD-SRSH spectra. For the bulk structure, we find that the bandstructure and optical spectrum is essen-tially insensitive to the choice of parameters, as may be expected given that both sets of values are optimal for bulk $MoS_2$ ($\Delta E_g = 0$). For the monolayer and bilayer, using the non-optimal set of parameters leads to nearly rigid shifts of the bandstructure by $\sim 0.5$ eV. The optical absorption spectra display lower sensitivity to this choice of parameters. For the bilayer, the only noteworthy change is in the amplitudes of the spectral features, whereas for the monolayer the differences in the energies of the spectral features ($\sim 0.1$ eV) is somewhat more noticeable. Thus, use of non-optimal parameters, $(\alpha^*, \gamma^*)$, for the bilayer will overestimate exciton binding energies by $\sim 0.5$ eV. Mitigating these errors may require a more complex multi-objective error function or the development of alternative dielectric screening models that explicitly account for the thickness of the 2D layer/slab [96–98].

### B. h-BN

The tuned SRSH parameters, $(\alpha^*, \gamma^*)$, for bulk and monolayer h-BN were reported in Ref. 37 and are listed in Table II. Following the same tuning procedure, we first perform GW calculations to determine the reference quasiparticle band gaps for h-BN bilayers (see SI and Table II). Figure 7(a,b) exhibit $\Delta E_g$ surfaces and contour plots for monolayer, bilayer, and bulk h-BN. Also in this case, the optimal parameters for the bilayer and bulk, labeled $\alpha'$ and $\gamma'$, are not identical to those optimized for the monolayer and bulk $(\alpha^*, \gamma^*)$, differing mostly in the range-separation parameter $\gamma$. The SRSH bandstructure for the bilayer is displayed in Figure 7(c), along with the reference GW calculation. Here the extrapolation difference is 160 meV. The two results are in good agreement. Considering the lowermost conduction band and the uppermost valence band, the mean absolute error is 0.207 eV. The TD-SRSH spectrum for the bilayer is displayed in Figure 7(c), along with the reference GW-BSE spectrum. Clearly, the former is blue-shifted by approximately 0.6 eV relative to latter. This is a known issue, discussed previously in Ref. 37 and is not pursued further here.

TABLE II. Brillouin zone grid, tuned SRSH parameters $(\alpha^*, \gamma^*)$ for bulk-monolayer, tuned SRSH parameters $(\alpha', \gamma')$ for bulk-bilayer, average inverse macroscopic dielectric constant $(\epsilon_\infty^{-1})$, GW band gap $(E^{GW})$, GW-BSE optical gap $(E_{opt}^{GW-BSE})$, SRSH band gap $(E^{SRSH}$, fitted to an extrapolated GW quasiparticle band gap) and TD-SRSH optical gap $(E_{opt}^{TD-SRSH})$ $MoS_2$ and h-BN. Additional computational details for the calculations are given in the SI. Band gaps and optical gaps are calculated at the K point for all phases.
[a]From Ref. 37.

| Material | Phase | k-grid | $\alpha^*$ | $\gamma^*(\text{Å}^{-1})$ | $\alpha'$ | $\gamma'(\text{Å}^{-1})$ | $\epsilon_\infty^{-1}$ | $E^{GW}$ [eV] | $E_{opt}^{GW-BSE}$ [eV] | $E^{SRSH}$ [eV] | $E_{opt}^{TD-SRSH}$ [eV] |
|---|---|---|---|---|---|---|---|---|---|---|---|
| **$MoS_2$** | Bulk | $12 \times 12 \times 4$[a] | 0.107 | 0.038 | 0.105 | 0.008 | 0.085 | 2.07[a] | 2.00[a] | 2.03[a] | 1.91 |
| | 2L | $15 \times 15 \times 1$ | | | | | 1.0 | 2.20 | 1.90 | 2.18 | 1.97 |
| | 1L | $18 \times 18 \times 1$[a] | | | | | 1.0 | 2.50[a] | 2.00[a] | 2.65[a] | 2.02 |
| **h-BN** | Bulk | $12 \times 12 \times 4$[a] | 0.201 | 0.072 | 0.204 | 0.041 | 0.25 | 6.58[a] | 5.48[a] | 6.66[a] | 5.82 |
| | 2L | $18 \times 18 \times 1$ | | | | | 1.0 | 6.79 | 5.29 | 6.95 | 5.91 |
| | 1L | $18 \times 18 \times 1$[a] | | | | | 1.0 | 7.20[a] | 5.31[a] | 7.26[a] | 5.92 |

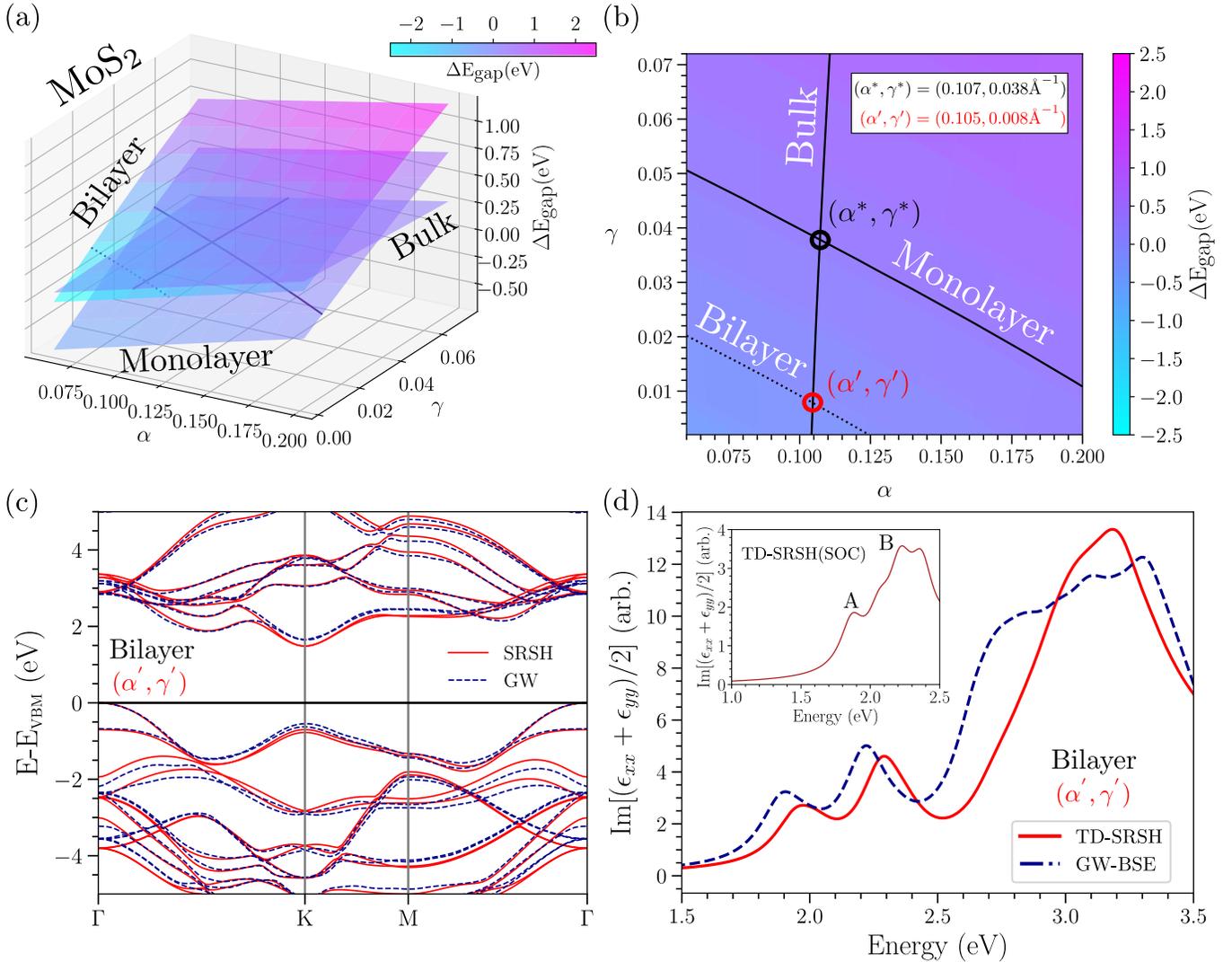

FIG. 5. (a) Surfaces and (b) 2D contour plots of the gap deviation, $\Delta E_g$, for monolayer, bilayer, and bulk $MoS_2$. Solid black lines represent values for which $\Delta E_g = 0$. The intersections of the solid lines yield a set of values $(\alpha^*, \gamma^*)$ that are transferable between the monolayer and bulk, and a somewhat different set of values $(\alpha', \gamma')$ that are transferable between the bilayer and bulk. (c) Bandstructures for bilayer $MoS_2$ from SRSH (red solid lines), using the parameters $(\alpha', \gamma')$, and from $G_0W_0$@PBE (blue dashed lines). (d) Optical absorption spectra for bilayer $MoS_2$ obtained with TD-SRSH (red solid line) and GW-BSE (blue dashed line). $\gamma$ has units of $\text{Å}^{-1}$.

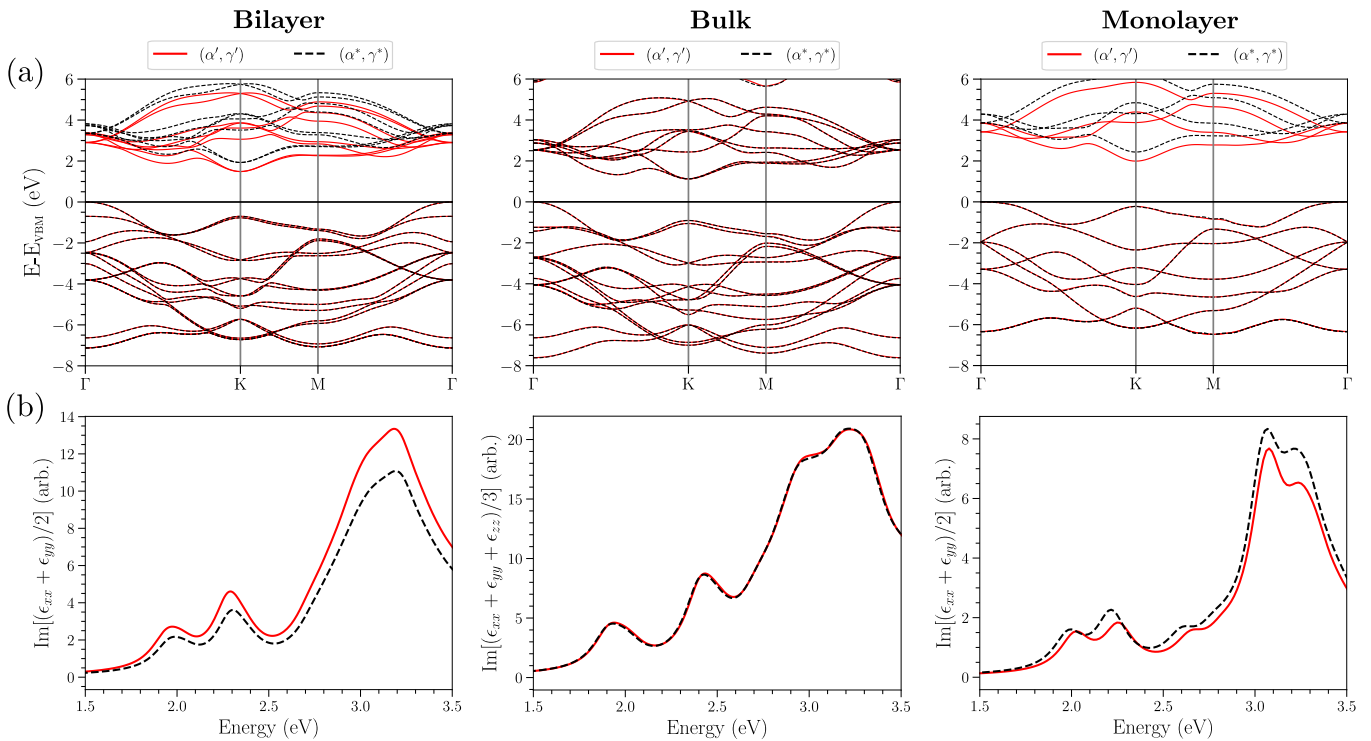

FIG. 6. (a) Band structures of MoS$_2$ in the bilayer, bulk and monolayer phases, calculated using the parameters indicated in Table II and with $(\alpha^*, \gamma^*) = (0.107, 0.038 \, \text{Å}^{-1})$ and $(\alpha', \gamma') = (0.105, 0.008 \, \text{Å}^{-1})$, obtained from the intersections between zero crossings indicated in Figure 5. The former pair is optimal for the monolayer and the latter for the bilayer. The choice of optimal versus non-optimal parameters leads to a near constant error of $\sim 0.5$ eV in band gaps along the indicated high-symmetry **k**-path for bilayer and monolayer. The band structure of the bulk is not sensitive to the choice of optimal parameters. (b) Optical absorption spectra for the phases and band structures in (a).

Finally, in Figure 8 we assess the transferability of the parameters $(\alpha^*, \gamma^*)$ and $(\alpha', \gamma')$ between monolayer, bilayer, and bulk h-BN, as done in Figure 6 for MoS$_2$. For bulk h-BN, the bandstructure and optical spectrum are insensitive to the choice of parameters. For the monolayer and bilayer, in contrast, using the non-optimal set of parameters leads to nearly rigid shifts of the bandstructure by $\sim 0.35$ eV. We also display in Figure 8 the TD-SRSH absorption spectra for the monolayer and bilayer using the two different sets of parameters for $\alpha$ and $\gamma$. Noting that the absorption spectrum suffers from a large blueshift, as discussed above, we only seek to understand *relative* differences between the TD-SRSH spectra. For both monolayer and bilayer h-BN, the position of the first excitonic peak changes only slightly by about 0.03 eV and the peak heights are also only slightly affected. Beyond the first peak though, the absorption spectra show more significant changes with the appearance of additional satellite peaks and/or shoulders. It is therefore likely that the overall SRSH/TD-SRSH functional formalism needs to be revisited for h-BN –and, possibly, other large-gap layered insulators.

## V. CONCLUSION

In conclusion, we have demonstrated a facile approach for the construction of transferable SRSH functionals for bulk and mono-/bilayer vdW materials. By tuning the SRSH functional to reproduce just one (GW) quasiparticle energy, we have demonstrated the ability to achieve excellent agreement between SRSH and GW bandstructures of bulk and mono-/bilayers TMDCs, black phosphorus, InSe, and h-BN, at a fraction of the computational cost of GW calculations. We have also shown that TD-SRSH calculations of excited-state properties, which do not enter at any stage into the functional tuning procedure, are generally in good agreement with the BSE approach, thus lending credence to the predictive capability of the SRSH/TD-SRSH formalism. The one exception to this finding is h-BN, the optical spectra of which are at variance with their BSE counterparts. As no such deviations have been reported before for SRSH studies of bulk insulators [35], it remains to be understood if this is a generic problem posed by large-gap 2D insulators and, if so, how to incorporate missing physical effects into the SRSH exchange-correlation kernel.

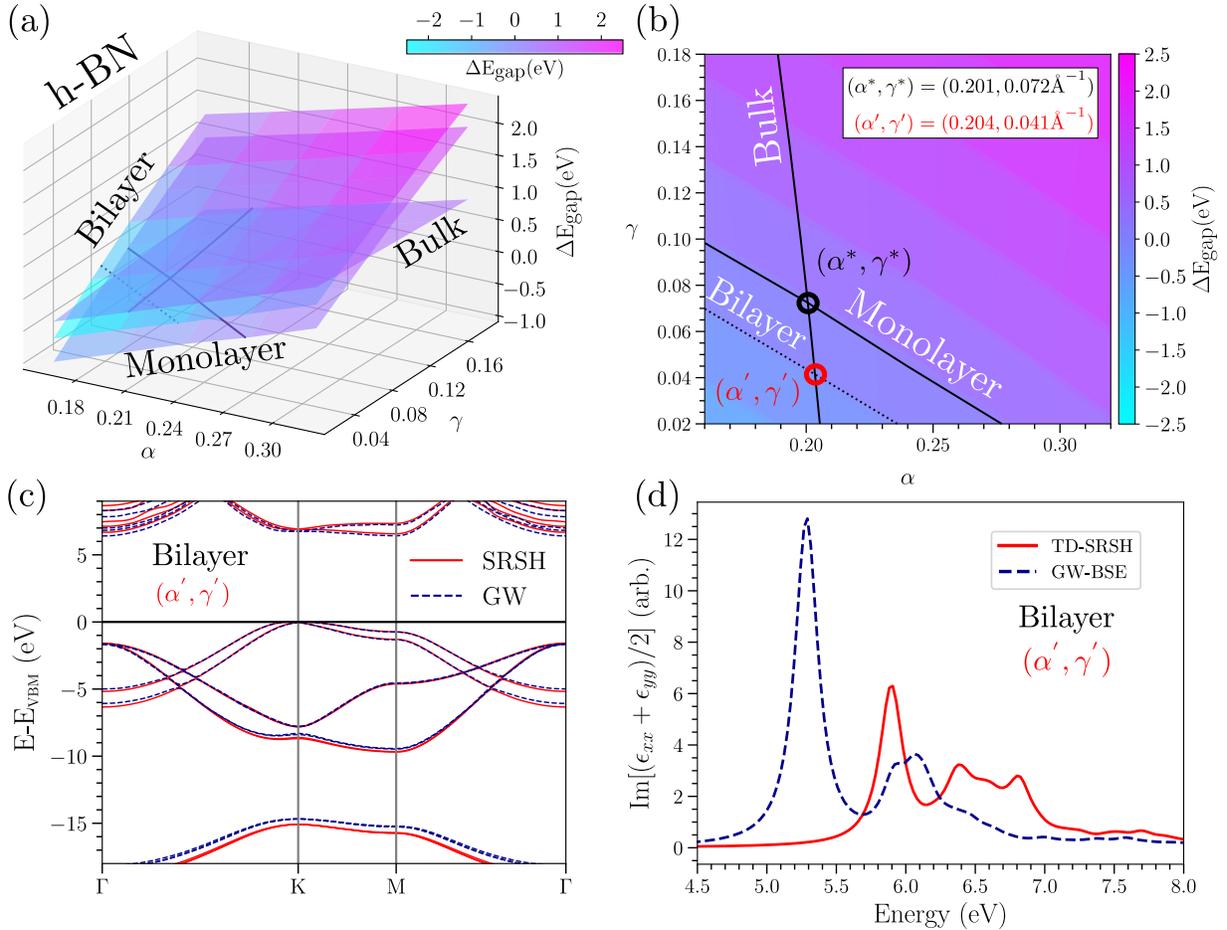

FIG. 7. (a) Surfaces and (b) 2D contour plots of the gap deviation, $\Delta E_g$, for monolayer, bilayer, and bulk h-BN. The solid black lines represent the values for which $\Delta E_g = 0$ for these structures. The intersections of the solid lines yield a set of values $(\alpha^*, \gamma^*)$ that are transferable between the monolayer and bulk, and another set of values $(\alpha', \gamma')$ that are transferable between the bilayer and bulk. (c) Bandstructures for bilayer h-BN from SRSH (solid lines) using parameters $(\alpha', \gamma')$ and $G_0W_0@PBE$ (dashed lines). (d) Optical absorption spectra for bilayer h-BN obtained with TD-SRSH (solid line) and GW-BSE (dashed line). $\gamma$ has units of Å$^{-1}$.

Our results suggest that the SRSH/TD-SRSH approach is robust for 2D semiconductors, opening up a range of opportunities for accurate calculations of the optoelectronic properties of layered materials with defects, hetero-layers/-junctions, twisted or shifted layers, and more.

## ACKNOWLEDGMENTS

This work was supported by the US Air Force through the grant AFOSR grant FA8655-20-1-7041, and by U.S.-Israel NSF–Binational Science Foundation Grant No. DMR-2015991. M.C.-G. is grateful to the Azrieli Foundation for the award of an Azrieli International Postdoctoral Fellowship. This work used Bridges2 [99] at Pittsburgh Supercomputing Center (PSC) through allocation TG-DMR190070 from the Extreme Science and Engineering Discovery Environment (XSEDE) [100], which was supported by National Science Foundation grant number #1548562. This work used Bridges2 at PSC through allocation TG-DMR190070 from the Advanced Cyberinfrastructure Coordination Ecosystem: Services & Support (ACCESS) program, which is supported by National Science Foundation grants #2138259, #2138286, #2138307, #2137603, and #2138296. Additional computational resources were provided by the

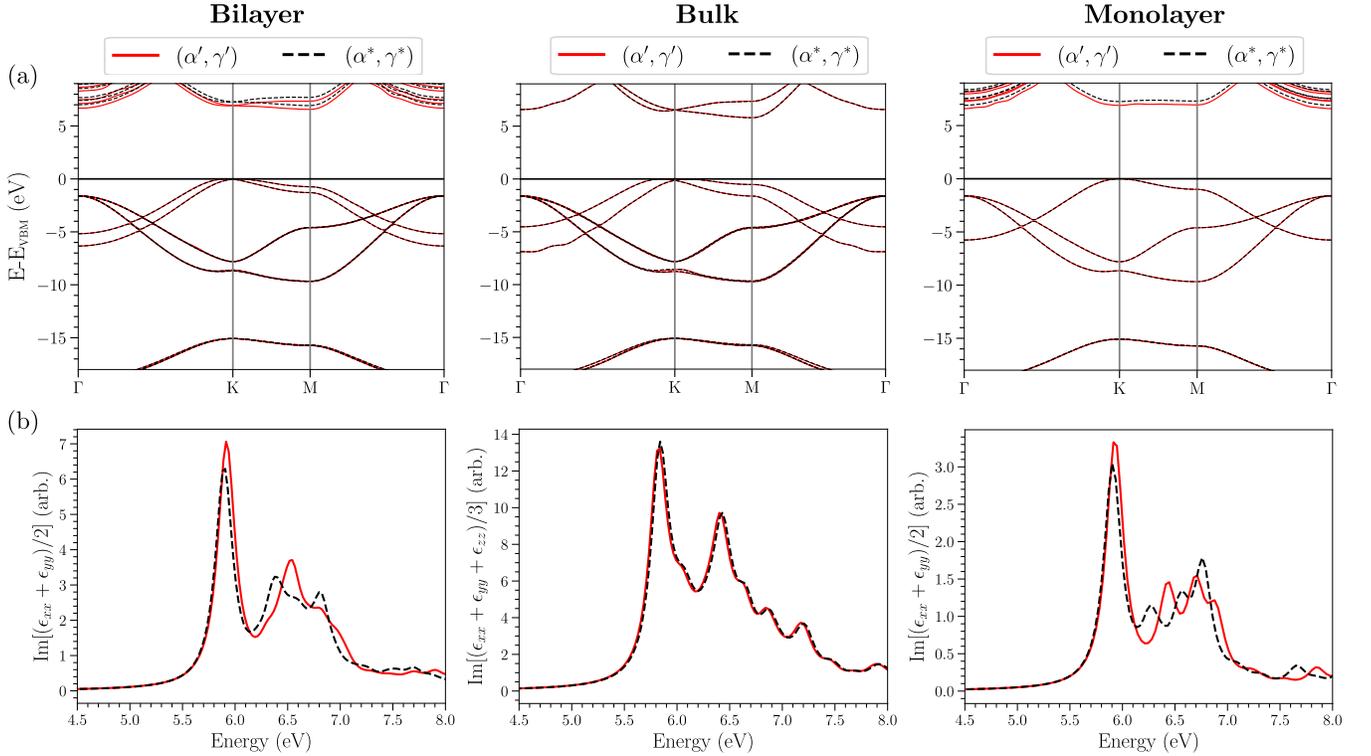

FIG. 8. (a) Band structures of h-BN in the bilayer, bulk and monolayer phases, calculated using the parameters indicated in Table II and with $(\alpha^*, \gamma^*) = (0.201, 0.072\,\text{Å}^{-1})$ and $(\alpha', \gamma') = (0.204, 0.041\,\text{Å}^{-1})$, obtained from the intersections between zero crossings indicated in Figure 5. The former pair is optimal for the monolayer and the latter for the bilayer. The choice of optimal versus non-optimal parameters leads to a near constant error of $\sim 0.35$ eV in band gaps along the indicated high-symmetry **k**-path for bilayer and monolayer. The band structure of the bulk is not sensitive to the choice of optimal parameters. (b) Optical absorption spectra for the phases and band structures in (a).

Weizmann Institute of Science at Chemfarm. L.K. acknowledges additional support from the Aryeh and Mintzi Katzman Professorial Chair and from the Helen and Martin Kimmel Award for Innovative Investigation.

<u>Supporting Information</u>

# Transferable screened range-separated hybrid functionals for electronic and optical properties of van der Waals materials

María Camarasa-Gómez,[1] Ashwin Ramasubramaniam,[2, 3]

Jeffrey B. Neaton,[4, 5, 6] and Leeor Kronik[1]

[1]*Department of Molecular Chemistry and Materials Science,*

*Weizmann Institute of Science, Rehovoth 7610001, Israel*

[2]*Department of Mechanical and Industrial Engineering,*

*University of Massachusetts Amherst, Amherst MA 01003, USA*

[3]*Materials Science and Engineering Graduate Program,*

*University of Massachusetts, Amherst, Amherst MA 01003, USA*

[4]*Department of Physics, University of California, Berkeley, CA 94720, USA*

[5]*Materials Sciences Division, Lawrence Berkeley*

*National Laboratory, Berkeley, CA 94720, USA*

[6]*Kavli Energy NanoSciences Institute at Berkeley,*

*University of California, Berkeley, CA 94720, USA*

**CONTENTS**

## S1. COMPUTATIONAL DETAILS

All results presented in this work were obtained using the Vienna *Ab Inito* Simulation Package[1,2] (VASP), v5.4.4. The GW flavor of the projector-augmented-wave pseudopotentials[3,4] (PAW) provided by VASP have been employed throughout, in both DFT and GW calculations. LDA-based PAWs were used for molybdenum disulfide (MoS$_2$) and hexagonal-boron nitride (h-BN), using an energy cutoff of 500 eV and 550 eV, respectively. PBE-based PAWs were used for the rest of the materials, employing energy cutoffs for the monolayer and the bulk phases, respectively, of 450 and 600 eV for molybdenum diselenide (MoSe$_2$), 600 eV for tungsten diselenide (WSe$_2$), 550 and 600 eV for tungsten disulfide (WS$_2$), 500 and 550 eV for indium selenide (InSe), and 400 and 450 eV for black phosphorus (BP). The electronic configuration for the valence electrons are $4s^2 4p^6 5s^1 4d^5$ for Mo, $3s^2 3p^4$ for S, $2s^2 2p^1$ for B, $2s^2 2p^3$ for N, $5p^6 6s^2 5d^4$ for W, $4s^2 4p^4$ for Se, $5s^2 5p^1$ for In, and $3s^2 3p^3$ for P.

The (semi-)local exchange-correlation used as ingredients in the SRSH calculations, as well as in the preparation of the wavefunctions employed in the GW calculations, has been the local density approximation[5,6] (LDA) for MoS$_2$ and h-BN. For BP, GW calculations were performed based on a DFT starting point calculated with the Heyd-Scuseria-Ernzerhof[7,8] (HSE) functional, and the Perdew-Burke-Ernzerhof[9] (PBE) approximation for the rest of materials presented in this work.

Benchmarks for the band structures and optical absorption spectra of the (time-dependent) screened range-separated hybrid functionals (TD-SRSH) are performed in a similar fashion to that reported in Ref. 10, employing the one-shot flavour of GW and the Bethe-Salpeter equation (BSE)[11–13]. Given the lack of Coulomb cutoff correction in VASP, the extrapolation of the band gap obtained with the $G_0 W_0$ benchmarks to an infinite vacuum distance is needed, and we report the selection of the band gap in Sec. S3. The macroscopic dielectric constants for the bulk systems were obtained for all materials using the Random Phase Approximation[14] (RPA) based on PBE, except for the case of BP, where it was calculated using the HSE functional. Spin-orbit coupling (SOC) has been taken into account only in TD-SRSH optical absorption spectra of monolayers for WS$_2$, WSe$_2$, MoSe$_2$, InSe monolayers, and the MoS$_2$ bilayer. All band structures were obtained through Wannier interpolation using Wannier90[15], v1.2.

All atomic structures employed in this work are given in Sec. S4 and were not relaxed. Only the bulk phase of BP was fully relaxed, with residual forces smaller than 0.01 eV/Å, with an energy tolerance smaller than $10^{-8}$ eV, and using Tkatchenko-Scheffler dispersion corrections[16].

## S2. GAP DEVIATION SURFACES FOR TRANSITION METAL DICHALCO-GENIDES

In this section we provide the gap deviation surfaces obtained following the procedure explained in Sec. 2 of the main text for the transition metal dichalcogenides $WS_2$, $WSe_2$ and $MoSe_2$, projections of which are shown in Fig. 1.

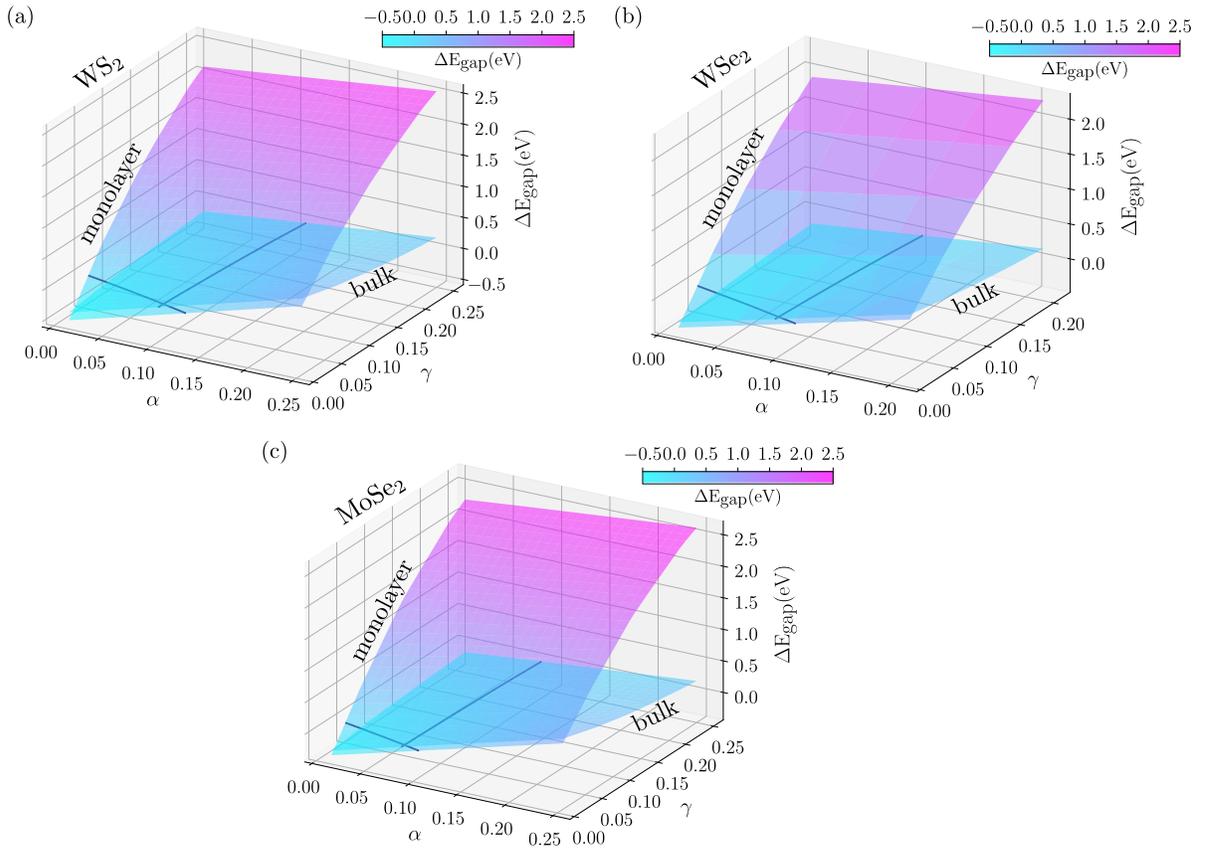

**Fig. S1**: Gap deviation surfaces, $\Delta E_{gap} = E_{gap}^{SRSH} - E_{gap}^{GW}$, obtained for (a) $WS_2$, (b) $WSe_2$, and (c) $MoSe_2$.

## S3. CONVERGENCE OF G₀W₀ CALCULATIONS

In this section we provide the extrapolation of the quasiparticle bandgaps and the selected bandgap fitting target for the SRSH calculations.

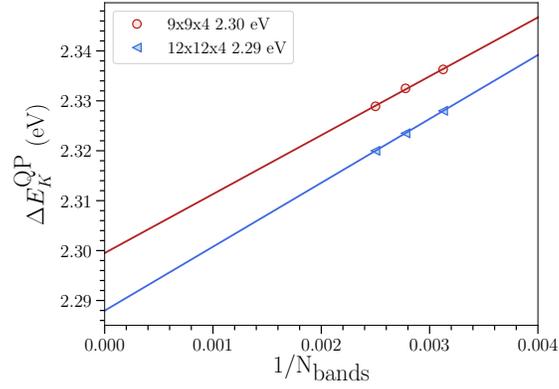

**Fig. S2**: Extrapolated $G_0W_0$@PBE quasiparticle bandgap, $E_K^{QP}$, at the $K$ point, in eV, for bulk $WS_2$, for different k-meshes as a function of the inverse number of bands, $1/N$. The legend also displays the extrapolated quasiparticle bandgap for each k-mesh. The extrapolated gap at the larger k-mesh has been used as the fitting target for the SRSH calculations.

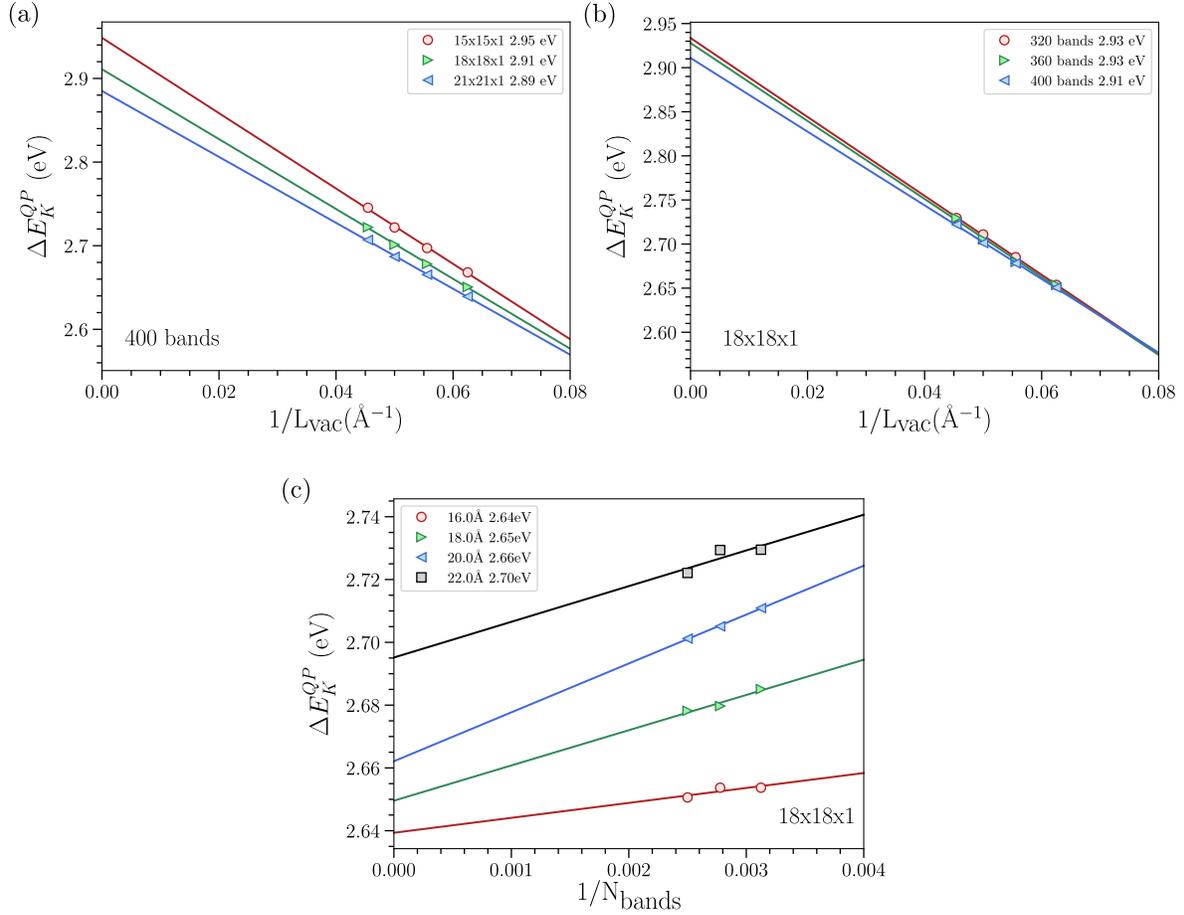

**Fig. S3**: Extrapolated $G_0W_0$@PBE quasiparticle bandgap, $E_K^{QP}$, at the $K$ point, in eV, for different k-meshes, for $WS_2$ monolayer (1L), as a function of the inverse vacuum distance for each k-mesh (a), different number of bands $1/N$ for a k-mesh (b), and different vacuum distances (c). The legend also displays the extrapolated quasiparticle bandgap. The extrapolated 2.66 eV with k-mesh $18 \times 18 \times 1$ is used as the target bandgap for the SRSH calculations.

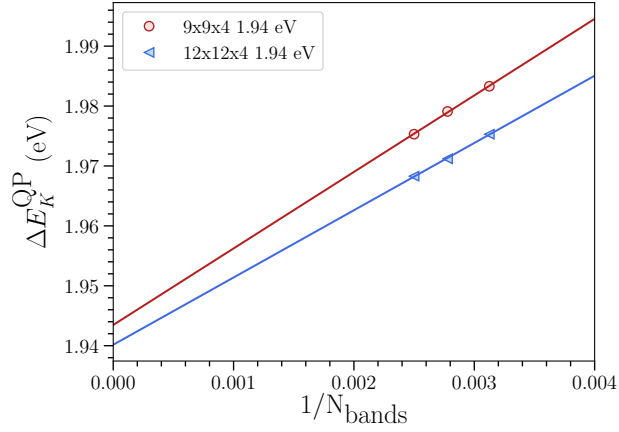

**Fig. S4**: Extrapolated $G_0W_0$@PBE quasiparticle bandgap, $E_K^{QP}$, at the $K$ point, in eV, for bulk $WSe_2$, for different k-meshes as a function of the inverse number of bands $1/N$. The legend also displays the extrapolated quasiparticle bandgap given for each k-mesh. The extrapolated gap at the larger k-mesh has been used as the fitting target for the SRSH calculations.

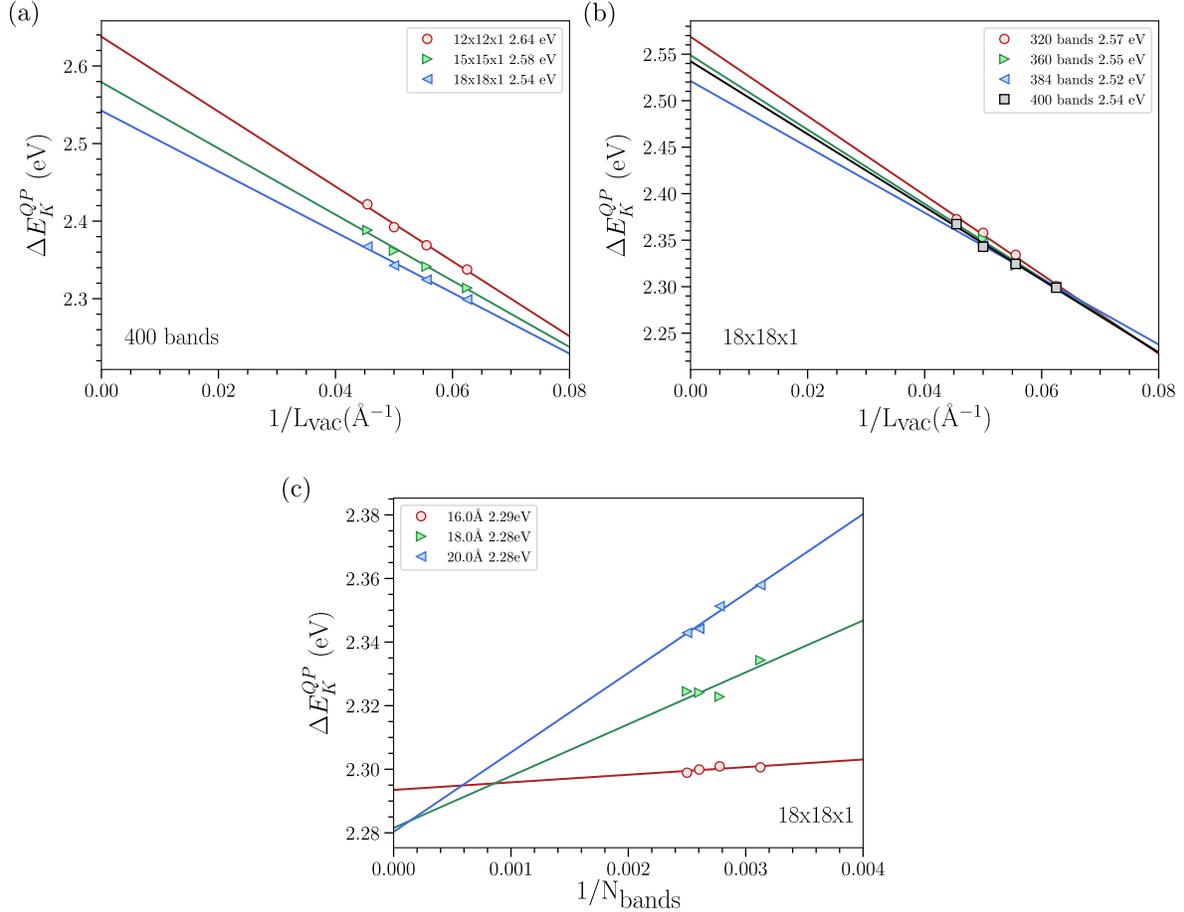

**Fig. S5**: Extrapolated $G_0W_0$@PBE quasiparticle bandgap, $E_K^{QP}$, at the $K$ point, in eV, for different k-meshes, for 1L $WSe_2$, as a function of the inverse vacuum distance for each k-mesh (a), different number of bands $1/N$ for a k-mesh (b), and different vacuum distances (c). The legend also displays the extrapolated quasiparticle bandgap. The extrapolated 2.28 eV with k-mesh $18 \times 18 \times 1$ is used as the target bandgap for the SRSH calculations.

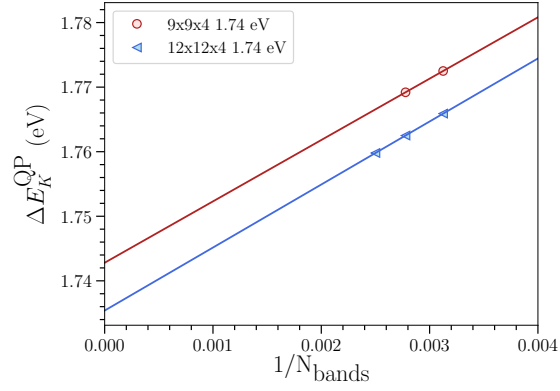

**Fig. S6**: Extrapolated $G_0W_0$@PBE quasiparticle bandgap, $E_K^{QP}$, at the $K$ point, in eV, for bulk MoSe$_2$, for different k-meshes as a function of the inverse number of bands $1/N$. The legend also displays the extrapolated quasiparticle bandgap given for each k-mesh. The extrapolated gap at the larger k-mesh has been used as the fitting target for the SRSH calculations.

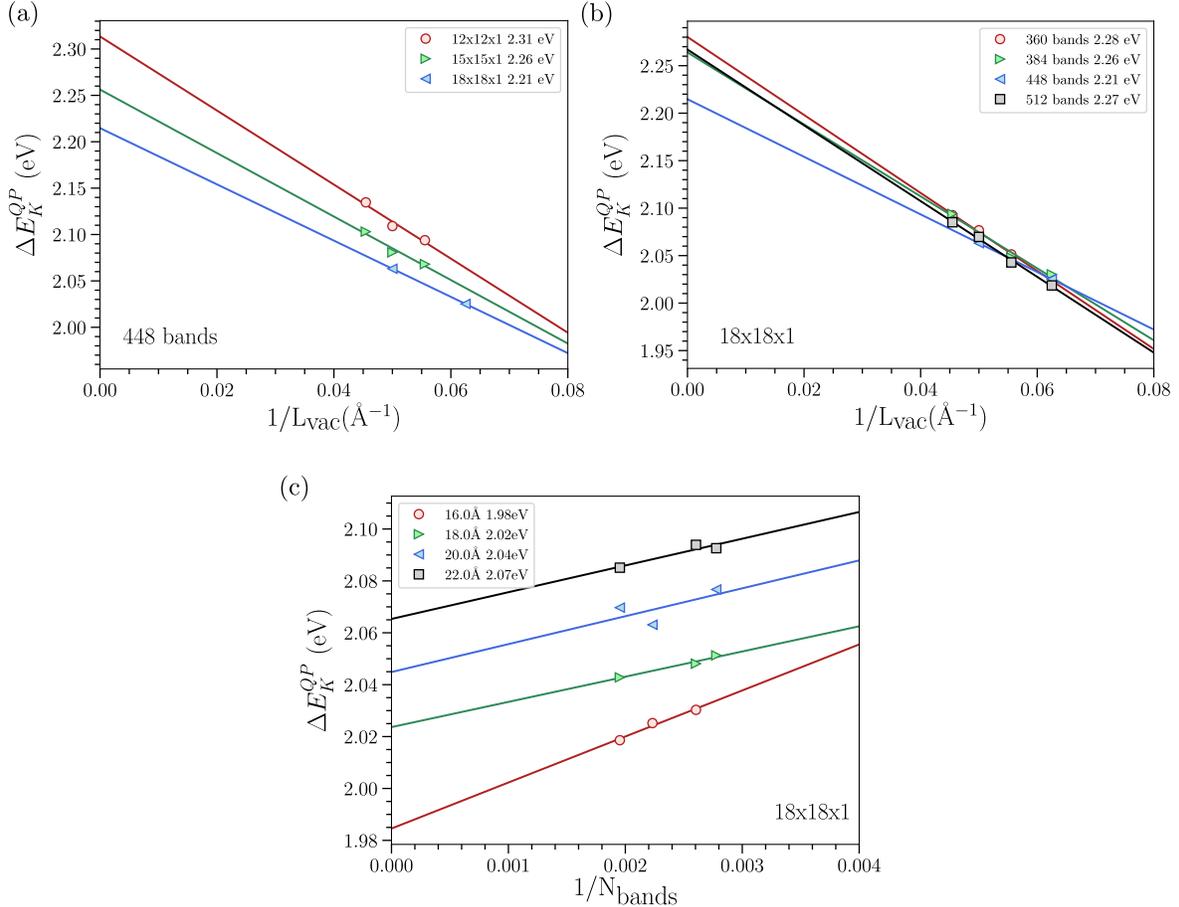

**Fig. S7**: Extrapolated $G_0W_0$@PBE quasiparticle bandgap, $E_K^{QP}$, at the $K$ point, in eV, for different k-meshes, for 1L MoSe$_2$, as a function of the inverse vacuum distance for each k-mesh (a), different number of bands $1/N$ for a k-mesh (b), and different vacuum distances (c). The legend also displays the extrapolated quasiparticle bandgap given for each k-mesh. The extrapolated 2.04 eV with k-mesh $18 \times 18 \times 1$ is used as the target bandgap for the SRSH calculations.

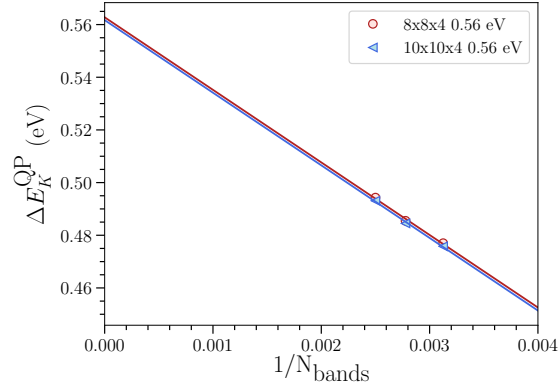

**Fig. S8**: Extrapolated $G_0W_0$@HSE quasiparticle bandgap, $E_K^{QP}$, at the $\Gamma$ point, in eV, for black phosphorus bulk, for different k-meshes as a function of the inverse number of bands $1/N$. The legend also displays the extrapolated quasiparticle bandgap given for each k-mesh. The extrapolated gap 0.56 eV at the k-mesh $8 \times 8 \times 4$ has been used as the fitting target for the SRSH calculations.

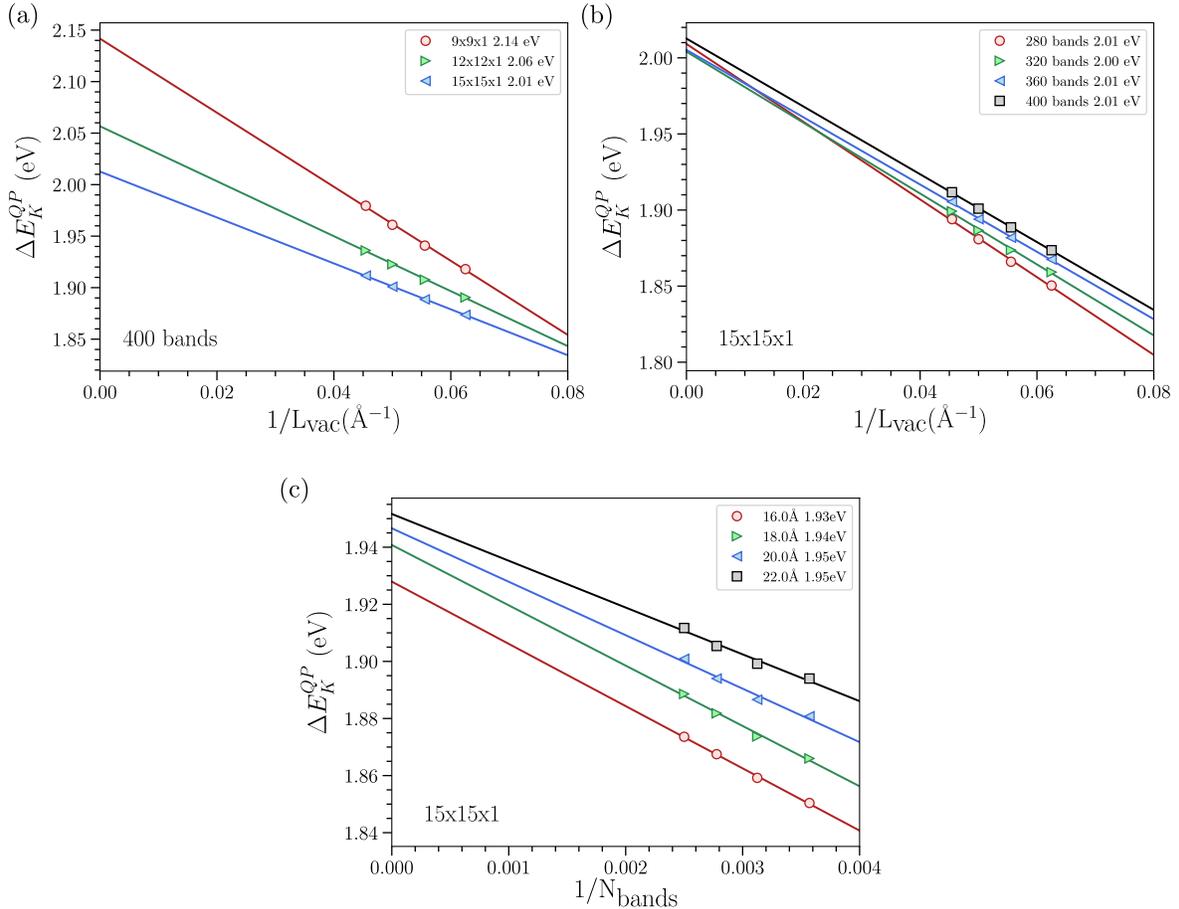

**Fig. S9**: Extrapolated $G_0W_0$@HSE quasiparticle bandgap, $E_K^{QP}$, at the $\Gamma$ point, in eV, for different k-meshes, for phosphorene, as a function of the inverse vacuum distance for each k-mesh (a), different number of bands $1/N$ for a k-mesh (b), and different vacuum distances (c). The legend also displays the extrapolated quasiparticle bandgap given for each k-mesh. The extrapolated 1.95 eV with k-mesh $15 \times 15 \times 1$ is used as the target bandgap for the SRSH calculations.

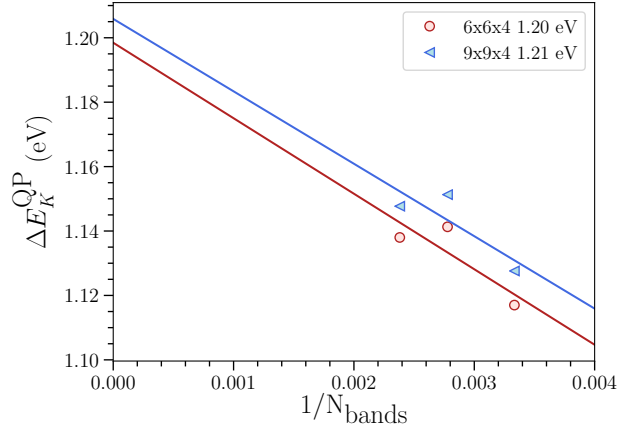

**Fig. S10**: Extrapolated $G_0W_0$@PBE quasiparticle bandgap, $E_K^{QP}$, at the $\Gamma$ point, in eV, for bulk InSe, for different k-meshes as a function of the inverse number of bands 1/N. The legend also displays the extrapolated quasiparticle bandgap given for each k-mesh. The extrapolated 1.21 eV with k-mesh $9 \times 9 \times 4$ is used as the target bandgap for the SRSH calculations.

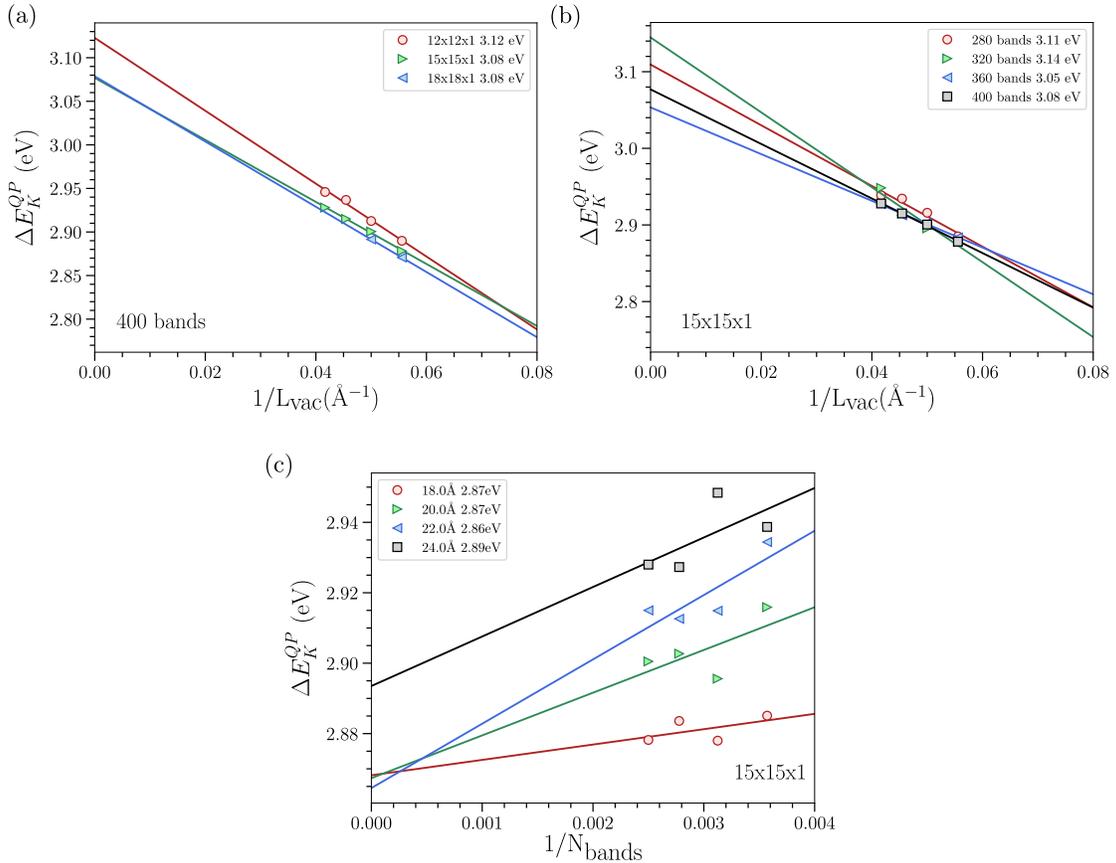

**Fig. S11**: Extrapolated $G_0W_0$@PBE quasiparticle bandgap, $E_K^{QP}$, at the $\Gamma$ point, in eV, for different k-meshes, for 1L InSe, as a function of the inverse vacuum distance for each k-mesh (a), different number of bands 1/N for a k-mesh (b), and different vacuum distances (c). The legend also displays the extrapolated quasiparticle bandgap. The extrapolated 2.87 eV with k-mesh $15 \times 15 \times 1$ is used as the target bandgap for the SRSH calculations.

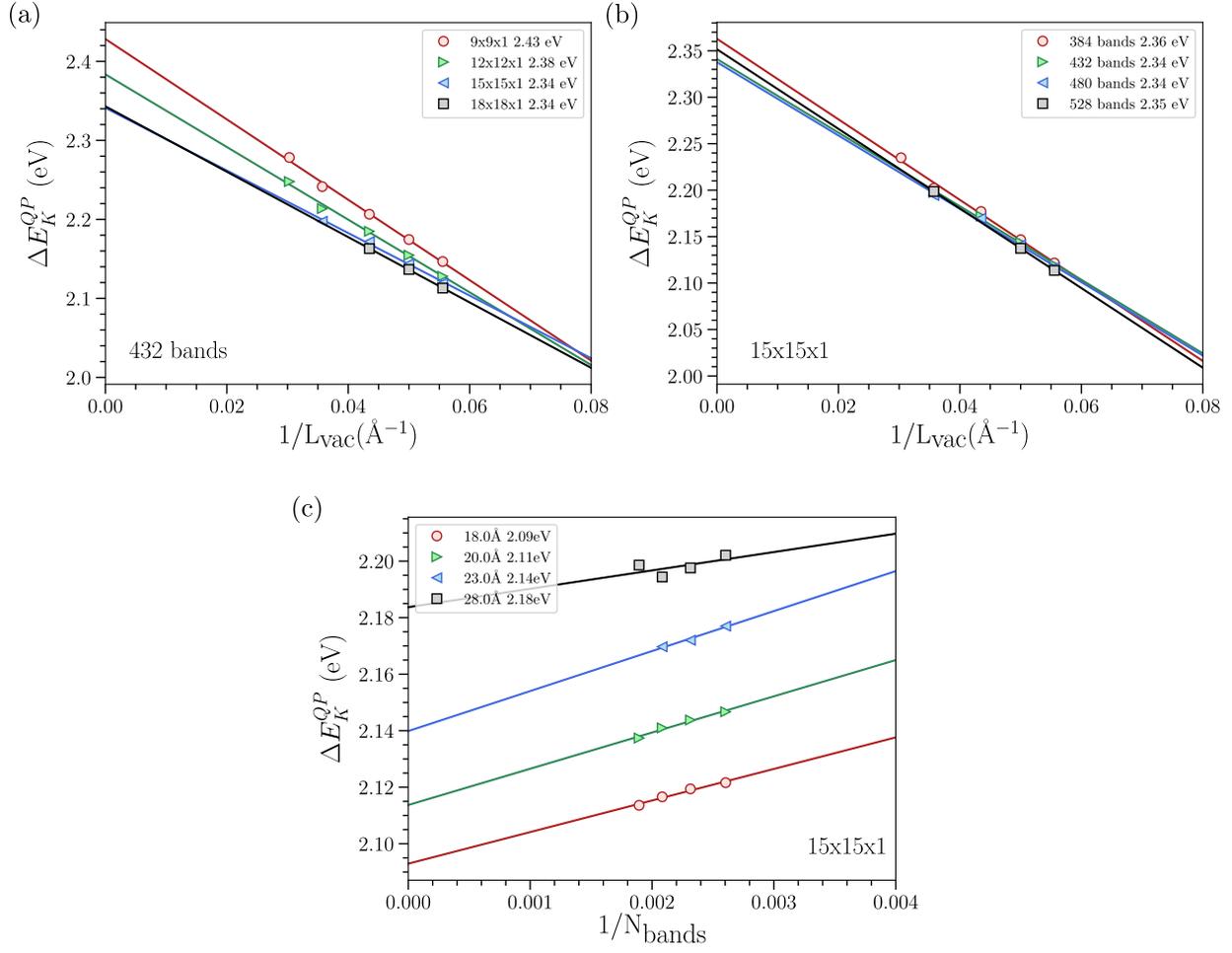

**Fig. S12**: Extrapolated $G_0W_0$@LDA quasiparticle bandgap, $E_K^{QP}$, at the $\Gamma$ point, in eV, for different k-meshes, for $MoS_2$ bilayer (2L), as a function of the inverse vacuum distance for each k-mesh (a), different number of bands $1/N$ for a k-mesh (b), and different vacuum distances (c). The legend also displays the extrapolated quasiparticle bandgap. The extrapolated 2.18 eV with k-mesh $15 \times 15 \times 1$ is used as the target bandgap for the SRSH calculations.

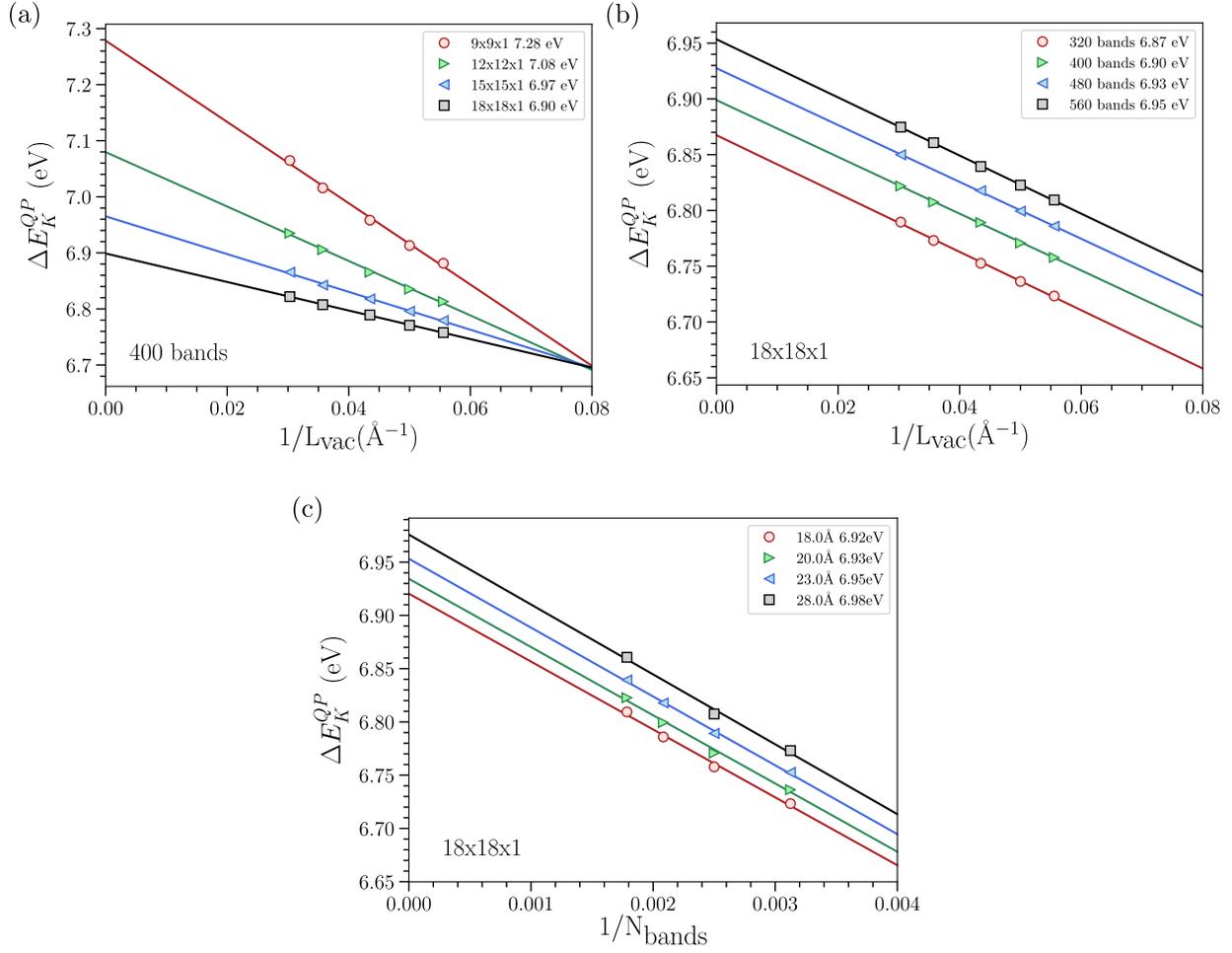

**Fig. S13**: Extrapolated $G_0W_0$@LDA quasiparticle bandgap, $E_K^{\mathrm{QP}}$, at the $\Gamma$ point, in eV, for different k-meshes, for 2L h-BN, as a function of the inverse vacuum distance for each k-mesh (a), different number of bands $1/N$ for a k-mesh (b), and different vacuum distances (c). The legend also displays the extrapolated quasiparticle bandgap. The extrapolated 6.95 eV with k-mesh $18 \times 18 \times 1$ is used as the target bandgap for the SRSH calculations.

## S4. ATOMIC GEOMETRIES

In this section, we provide the atomic structures of the bulk and monolayers in the format of VASP. The monolayer structures are obtained from the bulk by isolating a single layer and adding vacuum in the direction perpendicular to the plane of the monolayer.

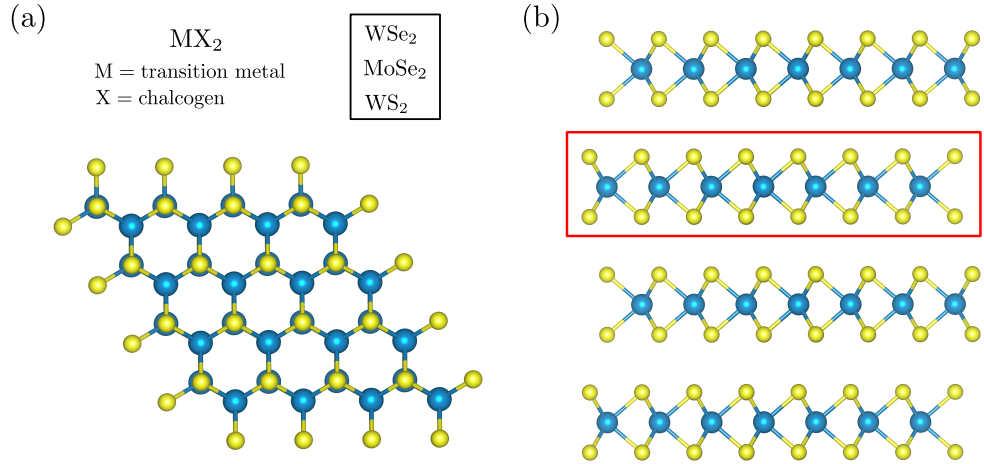

**Fig. S14**: Atomic structure for $WS_2$: (a) Top view and general chemical formula for transition metal dichalcogenides, and (b) lateral view. The red square contains the monolayer structure. $WSe_2$ and $MoSe_2$ present a similar structure.

**Unit cell of WS₂**

Atomic structure obtained from Ref. 17.

| **Bulk WS₂** | **Monolayer WS₂** |
|---|---|
| W S2 | W S2 |
| 1.0 | 1.0 |
| 3.1531999111 0.0000000000 0.0000000000 | 3.1531999111 0.0000000000 0.0000000000 |
| -1.5765999556 2.7307512262 0.0000000000 | -1.5765999556 2.7307512262 0.0000000000 |
| 0.0000000000 0.0000000000 12.3229999542 | 0.0000000000 0.0000000000 20.0 |
| W S | W S |
| 2 4 | 1 2 |
| Cartesian | Cartesian |
| 0.000000000 1.820500872 3.080749989 | 0.000000000 1.820500872 3.080749989 |
| 1.576599862 0.910250354 9.242249966 | 1.576599862 0.910250354 4.651932453 |
| 0.000000000 1.820500872 7.671067501 | 1.576599862 0.910250354 1.509567524 |
| 1.576599862 0.910250354 4.651932453 | |
| 1.576599862 0.910250354 1.509567524 | |
| 0.000000000 1.820500872 10.813432430 | |

**Unit cell of WSe₂**

Atomic structure obtained from Ref. 17.

**Bulk WSe₂**

W Se2

1.0

3.2820000648 0.0000000000 0.0000000000

-1.6410000324 2.8422954314 0.0000000000

0.0000000000 0.0000000000 12.9600000381

W Se

2 4

Cartesian

0.000000000 1.894863677 3.240000010

1.640999935 0.947431754 9.720000029

0.000000000 1.894863677 8.049456134

1.640999935 0.947431754 4.910543904

1.640999935 0.947431754 1.569456115

0.000000000 1.894863677 11.390543924

**Monolayer WSe₂**

W Se2

1.0

3.2820000648 0.0000000000 0.0000000000

-1.6410000324 2.8422954314 0.0000000000

0.0000000000 0.0000000000 20.0

W Se

1 2

Cartesian

0.000000000 1.894863677 3.240000010

1.640999935 0.947431754 4.910543904

1.640999935 0.947431754 1.569456115

**Unit cell for MoSe₂**

Atomic structure obtained from Ref. 18.

**Bulk MoSe₂**

Mo Se2

1.0

3.2850000858 0.0000000000 0.0000000000

-1.6425000429 2.8448935258 0.0000000000

0.0000000000 0.0000000000 12.9010000229

Mo Se

2 4

Cartesian

0.000000000 1.896595740 3.225250006

1.642499945 0.948297785 9.675750017

0.000000000 1.896595740 8.063125014

1.642499945 0.948297785 4.837875009

1.642499945 0.948297785 1.612625003

0.000000000 1.896595740 11.288375020

**Monolayer MoSe₂**

Mo Se2

1.0

3.2850000858 0.0000000000 0.0000000000

-1.6425000429 2.8448935258 0.0000000000

0.0000000000 0.0000000000 20.0

Mo Se

1 2

Cartesian

0.000000000 1.896595740 3.225250006

1.642499945 0.948297785 4.837875009

1.642499945 0.948297785 1.612625003

**Unit cell of black phosphorus**

The atomic structure was obtained from Ref. 19. Note that for this material alone, the bulk structure was relaxed, as noted before. The monolayer was then derived from the bulk structure without further structural relaxation.

**Bulk black phosphorus**

P

1.0

3.3132998943 0.0000000000 0.0000000000

0.0000000000 10.4729995728 0.0000000000

0.0000000000 0.0000000000 4.3740000725

P

Cartesian

0.000000000 1.082908150 0.352544410

0.000000000 9.390091501 4.021455565

0.000000000 4.153591714 2.539544544

0.000000000 6.319407858 1.834455659

1.656649947 6.319407858 0.352544410

1.656649947 4.153591714 4.021455565

1.656649947 9.390091501 2.539544544

1.656649947 1.082908150 1.834455659

**Monolayer black phosphorus**

P

1.0

3.3149049282 0.0000000000 0.0000000000

0.0000000000 4.4248895645 0.0000000000

0.0000000000 0.0000000000 20.0

P

Cartesian

0.000000000 2.575827533 2.143897536

0.000000000 1.849062296 0.000000000

1.657452464 0.363382520 0.000000000

1.657452464 4.061507078 2.143897536

(a)                                    (b)

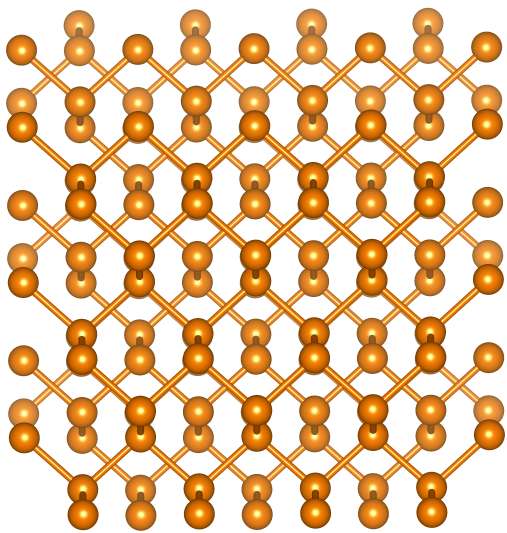   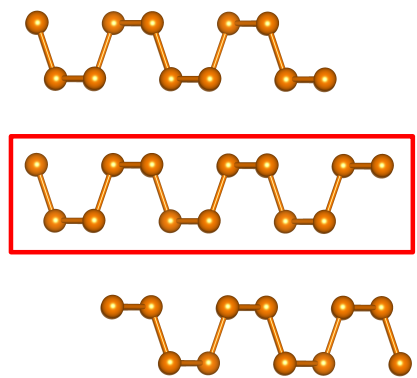

**Fig. S15**: Atomic structure for black phosphorus (BP): (a) Top view, and (b) lateral view. The red square contains the monolayer structure.

## Unit cell of InSe

The atomic structure was obtained from Ref. 20.

**Bulk InSe**

In Se

1.0

4.0036997795 0.0000000000 0.0000000000

-2.0018498898 3.4673057182 0.0000000000

0.0000000000 0.0000000000 16.6439990997

In Se

4 4

Cartesian

0.000000000 2.311537214 2.794527383

2.001849770 1.155768504 13.849472213

2.001849770 1.155768504 11.116526437

0.000000000 2.311537214 5.527472167

0.000000000 2.311537214 15.146038625

2.001849770 1.155768504 1.497959978

2.001849770 1.155768504 6.824039571

0.000000000 2.311537214 9.819960024

**Monolayer InSe**

In Se

1.0

4.0036997795 0.0000000000 0.0000000000

-2.0018498898 3.4673057182 0.0000000000

0.0000000000 0.0000000000 20.0

In Se

2 2

Cartesian

0.000000000 2.311537214 2.794527383

0.000000000 2.311537214 5.527472167

2.001849770 1.155768504 1.497959978

2.001849770 1.155768504 6.824039571

(a)

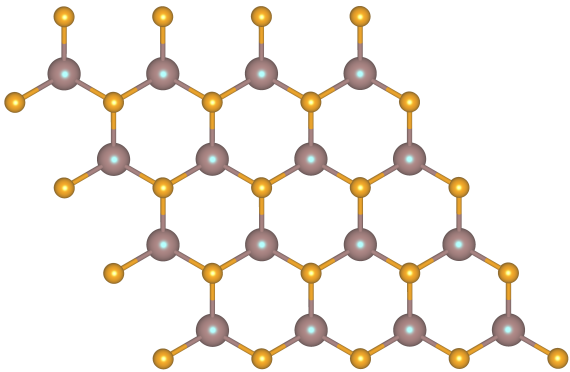

(b)

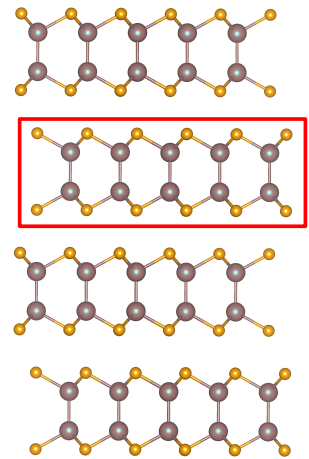

**Fig. S16**: Atomic structure for InSe: (a) Top view, and (b) lateral view. The red square contains the monolayer structure.

**Bilayer MoS₂**

The atomic structure was obtained from Ref. 21.

Mo S2

1.0

3.1610000134 0.0000000000 0.0000000000

-1.5805000067 2.7375063129 0.0000000000

0.0000000000 0.0000000000 28.0

Mo S

2 4

Cartesian

0.000000000 1.825004263 3.073750019

1.580499912 0.912502050 9.221250057

0.000000000 1.825004263 7.715112519

1.580499912 0.912502050 4.579887558

1.580499912 0.912502050 1.567612480

0.000000000 1.825004263 10.727387596

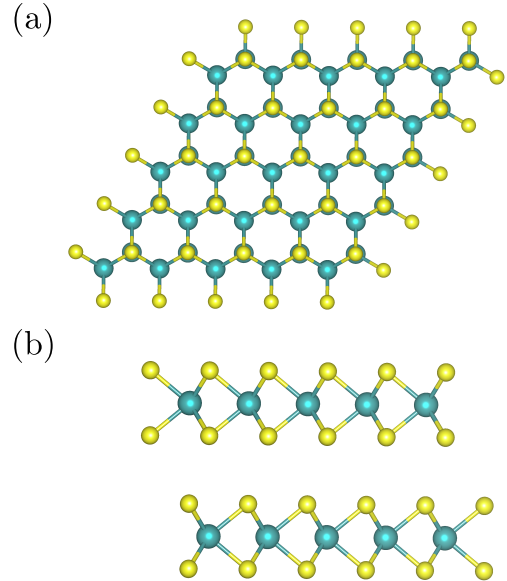

(a)

(b)

**Fig. S17**: Atomic structure for bilayer MoS₂: (a) Top view, and (b) lateral view.

**Bilayer h-BN**

The atomic structure was obtained from Ref. 22.

BN

1.0

2.4982399940 0.0000000000 0.0000000000

-1.2491199970 2.1635392996 0.0000000000

0.0000000000 0.0000000000 23.0

B N

2 2

Cartesian

0.000000000 0.000000000 3.317850113

0.000000000 1.442359576 0.000000000

0.000000000 0.000000000 0.000000000

0.000000000 1.442359576 3.31785011

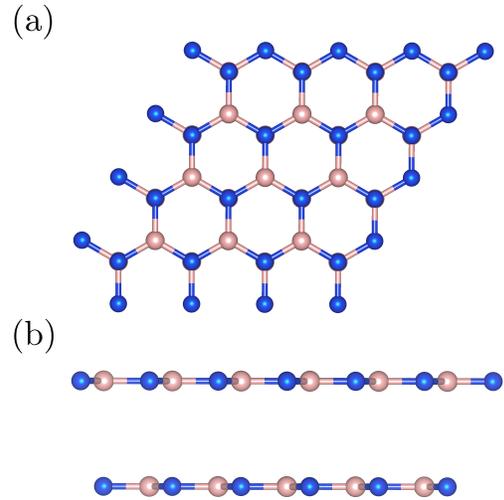

**Fig. S18**: Atomic structure for bilayer h-BN: (a) Top view, and (b) lateral view.